\documentclass[final,1p,times]{elsarticle}

\usepackage{latexsym,mathrsfs}
\usepackage{amsthm,enumerate,verbatim}
\usepackage{amsfonts}
\usepackage{url}
\usepackage[utf8]{inputenc}
\usepackage{hyperref}
\usepackage{caption}
\usepackage{subcaption}
\usepackage{amsmath}
\usepackage{bbm}
\usepackage{cases}
\usepackage{amssymb}
\usepackage{paralist}
\usepackage{multirow}
\usepackage{makecell}
\usepackage{booktabs}
\usepackage{siunitx}
\usepackage[table]{xcolor}
\usepackage{array}
\usepackage{graphicx}
\usepackage{threeparttable}
\usepackage{algorithm}
\usepackage{algcompatible}
\usepackage{float}
\usepackage{comment}
\usepackage{wrapfig}

\newtheorem*{definition*}{Definition}

\graphicspath{{figures/}}

\definecolor{brightpink}{rgb}{1.0, 0.0, 0.5}

\journal{Computer Networks}

\begin{document}

\begin{frontmatter}



\title{Structured Nonnegative Matrix Factorization for Traffic Flow Estimation of Large Cloud Networks}


\author[add:gsse-kiet]{Syed Muhammad Atif}
\ead{s.m.atif@kiet.edu.pk.}
\author[add:mor-uni-mons]{Nicolas Gillis\corref{ca:ng}}
\cortext[ca:ng]{Corresponding author: (Nicolas Gillis) Email: nicolas.gillis@umons.ac.be., Tel.: +32-(0)65-374680, Fax: +32-(0)65-374500;}
\ead{nicolas.gillis@umons.ac.be.}
\ead[url]{https://sites.google.com/site/nicolasgillis/}
\author[add:coe-kiet]{Sameer Qazi}
\ead{sameer.qazi@pafkiet.edu.pk.}
\author[add:coe-kiet,add:seece-uwa]{Imran Naseem}
\ead{imrannaseem@pafkiet.edu.pk,imran.naseem@uwa.edu.au.}

\address[add:gsse-kiet]{{Graduate School of Science and Engineering, Karachi Institute of Economics and Technology},
            {Korangi Creek}, 
            {Karachi},
            {75190}, 
            {Sindh}, 
            {Pakistan}}
            
\address[add:coe-kiet]{{College of Engineering, Karachi Institute of Economics and Technology},
            {Korangi Creek}, 
            {Karachi},
            {75190}, 
            {Sindh},
            {Pakistan}}
\address[add:mor-uni-mons]{{Department of Mathematics and Operational Research Faculté polytechnique, Université de Mons},
            {Rue de Houdain 9},
            {7000}, 
            {Mons},
            {Belgium}}
\address[add:seece-uwa]{{School of Electrical, Electronic and Computer Engineering, The University of Western Australia},
            {35 Stirling Highway}, 
            {Crawley},
            {6009}, 
            {Western Australia},
            {Australia}}

\begin{abstract}
		Network traffic matrix estimation is an ill-posed linear inverse problem: it requires to estimate the unobservable origin destination traffic flows, $X$,   given the observable link traffic flows,   $Y$, and a binary routing matrix, $A$, which are such that $Y= AX$. 
		This is a challenging but vital problem as accurate estimation of OD flows is required for several network management tasks. 
		In this paper, we propose a novel model for the network traffic matrix estimation problem which maps high-dimension OD flows to low-dimension latent flows with the following three constraints:
		\begin{inparaenum}[(1)]
			\item nonnegativity constraint on the estimated OD flows, 
			\item autoregression constraint that enables the proposed model to effectively capture temporal patterns of the OD flows, and 
			\item orthogonality constraint that ensures the mapping between low-dimensional latent flows and the corresponding link flows to be distance preserving. 
		\end{inparaenum}		
		The parameters of the proposed model are estimated with a training algorithm based on Nesterov accelerated gradient and generally
		shows fast convergence. 
		 We validate the proposed traffic flow estimation model on two real backbone IP network datasets, namely  Internet2 and G\'EANT. Empirical results show that the proposed model outperforms the state-of-the-art models not only in terms of tracking the individual OD flows but also in terms of standard performance metrics. 
		 The proposed model is also found to be highly scalable compared to the existing state-of-the-art approaches. 

\end{abstract}



		
		




\end{frontmatter}


    	\section{Introduction} \label{sec_intro}
    
    	Since the emergence of Internet, network parameter estimation has emerged to be an important but challenging research topic whose goal is to infer the network parameters that {cannot be} measured, such as end-to-end traffic volume and link delays. 
    	The significance of this research domain is due to its vital role in the network engineering tasks such as capacity planning~\cite{xie_accurate_2018}, network optimization~\cite{xiao_anomaly-tolerant_2019}, expansion and monitoring of the network~\cite{hu_accurate_2008,tune_fisher_2011}, congestion avoidance~\cite{pan_identify_2019}, and anomaly detection~\cite{li_online_2019}. In this paper, we focus on the network traffic matrix estimation {problem that estimates the origin-to-destination traffic volumes} using readily available routing matrix and easily observable link traffic volume.

    	Consider a network which consists of $m$ links and $n$ origin-destination (OD) pairs. Let the network be observed at time $t$, and let us denote $\boldsymbol{x}_{t} \in \mathbb{R}_{+}^{n}$ and $\boldsymbol{y}_{t} \in \mathbb{R}_{+}^{m}$ the OD flow and link flow vectors at time $t$, respectively. Further, let $A \in \{0,1\}^{m \times n}$ be the {binary} routing matrix of the network such that $A(i,j) = 1$ if the routing path for the $j$th OD pair of the network  passes through the $i$th link of the network. Then, the following relationship between link flow vector $\boldsymbol{y}_{t}$ and OD flow vector $\boldsymbol{x}_{t}$ holds: 
    	\begin{equation}\label{eq:traffic_tomography_column}
    		\boldsymbol{y}_{t} = A\boldsymbol{x}_{t} . 
    	\end{equation}
    	When the network is monitored for a time period $T$,  \eqref{eq:traffic_tomography_column} becomes:
    	\begin{equation}\label{eq:traffic_tomography_matrix}
    		Y = AX
    	\end{equation}     
    	where the link flow matrix $Y \in \mathbb R_{+}^{m \times T}$ and traffic matrix $X \in \mathbb R_{+}^{n \times T}$ are formed by stacking the column vectors $\boldsymbol{y}_{t}$'s and $\boldsymbol{x}_{t}$'s, for $t = 1,2,\dots,T$. 
    	
    	The traffic estimation problem is the {linear} inverse problem of recovering $X$ from $A$ and $Y$ using~\eqref{eq:traffic_tomography_matrix}. More precisely, the goal is to estimate the OD flows $\boldsymbol{x}_{T+i}$ at the future timestamps $T+i$ (for $i =1,2,\dots$) provided that the corresponding link flows $\boldsymbol{y}_{T+i}$, the binary routing matrix $A$, and the historical OD flows $X \in \mathbb R_{+}^{n \times T}$  and link flows measurements $Y \in \mathbb R_{+}^{m \times T}$ are given. 
        It is an ill-posed problem for all practical networks because the number of links in the networks is typically much smaller than the OD pairs in the network, that is, $m \ll n$. 
    	Hence, finding a solution for this problem is challenging because of the following two reasons: 
    	\begin{inparaenum}[(1)]
    	
    		\item Due to the rapid advancement in the network technology and its abundance at an affordable price, the size of modern networks are growing day by day~\cite{tune_internet_2013}. Hence, the difference between the number of links $m$ and that of OD pairs $n$ is becoming more and more prominent.  
    		
    		\item Due to the increase in diversity of applications running on the modern Internet, the statistical features of traffic are being further complicated~\cite{wu_growth_2011}.
    		
    	\end{inparaenum}

    	Dimensionality reduction techniques play vital rule in network traffic estimation problem. It is because OD flows when observed over a period of time $T$, that is, the OD flow matrix $X \in \mathbb{R}_{+}^{n \times T}$, have diurnal cycles hence high-dimensional OD flows can be mapped to low-dimensional latent flows using a dimensionality reduction technique such as the truncated SVD. 
    	Let us consider the following 
    	low-rank approximation 
    	\begin{equation}\label{eq:lrf_traffic_estimation}
    		X \approx PQ,
    	\end{equation}   
    	where $P \in \mathbb{R}^{n \times k}$ and $Q \in \mathbb{R}^{k \times T}$ are the left and right low-rank factors, and $k < m \ll n$ is the factorization rank. The matrix $P$ encodes the spatial features of OD flows whereas $Q$ encodes the latent flows associated with the OD flow matrix $X$. 
    	Plugging \eqref{eq:lrf_traffic_estimation} in \eqref{eq:traffic_tomography_matrix}, we obtain 
    	\begin{equation}\label{eq:traffic_tomography_using_lrc}
    		Y \approx APQ. 
    	\end{equation}
    	For a backbone Internet network, it can be safely assumed that the routing matrix $A$ and spatial feature matrix $P$ remain stable over a long period of time. Thus, the highly ill-posed problem of estimating OD flows using \eqref{eq:traffic_tomography_matrix} is transformed with the help of dimensionality reduction into the well-posed problem of estimating latent flows $Q$ using \eqref{eq:traffic_tomography_using_lrc}, given that $P$ has been estimated using historical data.   
    	In fact, given a new link flow observation $\boldsymbol{y}_t$, we can estimate $\boldsymbol{x}_{t}$ by solving  $\min_{\boldsymbol{q}_t \in \mathbb{R}^k} \| \boldsymbol{y}_t - AP\boldsymbol{q}_t \|_2$ whose solution is given by $\boldsymbol{q}_t = (AP)^{\dagger} \boldsymbol{y}_t$ where $(AP)^{\dagger}$ is the 
    	(left) pseudoinverse of $AP \in \mathbb{R}^{m \times k}$, and obtain $\boldsymbol{x}_{t} = P \boldsymbol{q}_t$. 

    	\subsection{Related work}\label{subsec:related_work}
    	
    	Several approaches have been proposed recently for traffic matrix estimation.
    	Among them dimensionality reduction techniques are the most popular and they are the focus of this paper. The reason why dimensionality reduction techniques can be employed in network traffic estimation is because 
    		\begin{inparaenum}[(1)]
    			\item the backbone Internet traffic is highly concentrated, that is,   most of the OD flows use only few links of the network, and 
    			\item OD flows have spatial and temporal similarities, that is, nearby OD pairs have similar traffic and OD flows have diurnal cycles.	
    		\end{inparaenum} 
    	Soule~et al.~\cite{soule_traffic_2005} proposed to employ the singular value decomposition (SVD)~\cite{golub_matrix_2013}, a well-known unconstrained dimensionality reduction technique, to develop a traffic estimation model. A disadvantage of their model is that it uses the 
    	Moore-Penrose pseudoinverse hence requiring an additional step to suppress possible negative OD flows in the initial estimate of the model. 
    	In~\cite{kumar_compressive_2018}, Kumar~et al.\ proposed a traffic estimation model based on a relatively new unconstrained dimensionality reduction technique known as CUR~\cite{mahoney_cur_2009}. Its key benefits over SVD are mainly computational efficiency and the interpretability of the low-rank factors that are directly derived from the given data. 
    	Recently, Qazi~et al.~\cite{qazi_novel_2018} proposed to use the demand matrix and the traffic probability matrix pair to transform the ill-posed traffic estimation problem into an equivalent well-posed problem. However, the two proposed methods~\cite{kumar_compressive_2018,qazi_novel_2018} suffer from the limitations similar to that of~\cite{soule_traffic_2005} due to the use of the Moore-Penrose pseudoinverse. 
    	
    	Some researchers have shown keen interest in different neural network architectures to design an effective model for traffic flow estimation. For example, Jiang~et al.\ in~\cite{jiang_accurate_2011} proposed to employ feed forward neural network (FFNN) for traffic flow estimation modeling. The proposed model BPTME takes link flows as input to FFNN whereas OD flows are yielded as output by the network. The network is trained by back error propagation algorithm. Zhou~et al.\ in~\cite{zhou_traffic_2016} proposed an enhancement over BPTME~\cite{jiang_accurate_2011} by injecting the routing information into the FFNN as input for improved performance. Due to the architectural simplicity, FFNN does not scale well with size, complexity and dynamics of traffic flows in the modern networks, hence Nie et al~\cite{nie_traffic_2016} proposed to a traffic flow estimation model based on deep belief network (DBN). In~\cite{zhao_towards_2018}, Zhao~et al.\ used long short term memory recurrent neural network (LSTM-RNN). DLTMP~\cite{zhao_towards_2018} is found to perform better compared to the contemporary approaches in capturing spatiotemporal dependencies of traffic flows because LSTM-RNN has cyclic connections over time.
    	
    	Genetic or evolutionary algorithms are also utilized by some researchers in the context of traffic flow estimation~\cite{zhang_network_2016,zhang_network_2016-1}. These typically employ quantum-behaved particle swarm optimization (QPSO). In~\cite{song_network_2019}, Lu~et al.\ used multi-fractal discrete wavelet transform (MDWT) to split traffic matrix into different frequency component then train the neural network to predict low and high frequency component of traffic matrix. Kumar~et al.\ in~\cite{kumar2020multi} proposed a multi-view subspace learning technique for traffic flow estimation. They proposed a novel robust approach to obtain traffic flows from multiple traffic views yielded from rather inexpensive existing methods.
    	
    	In this paper, we consider graph embedding and nonnegative matrix factorization (NMF) to perform traffic flow  estimation. 
    	Graph embedding has been successfully applied in various research fields such as computer vision and recommender systems. Interested readers may refer to the recent survey paper~\cite{cai2018comprehensive} by Cai~et al.\ on graph embedding for further details. Emami~et al.~\cite{emami2019new} have recently used graph embedding in the context of traffic flow estimation. Their approach blends graph embedding with convolution neural network (CNN) for better traffic flow estimation. 
    	Moreover, graph embedding has also been used by several researchers in combination with matrix factorization. A seminal research work in the context of NMF along with graph embedding is~\cite{deng_cai_graph_2011} by Cai~et al. 
    	Roughan~et al.~\cite{roughan_spatio-temporal_2012} have used classical NMF in the scenario of network traffic flow estimation. An autoregression based approach has been proposed recently for capturing temporal dependencies in~\cite{yu_temporal_2016} 
    	by Yu~et al.  This autoregressive approach maps the high-dimensional time series data into a low-dimensional latent time series. To avoid overfitting, each latent timeseries is autoregressed independently. Further, the authors of~\cite{yu_temporal_2016} proved that their proposed autoregressive approach has an equivalent graph representation, thus conventional solvers used for graph embedding approaches can be used for this autoregression approach as well. 
	    
	    \subsection{Contribution and outline of the paper}
	
	    Motivated from the benefits of NMF and autoregression approaches, and keeping in mind the limitations of the existing traffic flow estimation models, we propose a novel model for traffic flow estimation that is developed using a unique  combination of NMF, autoregression and orthogonality. 
 
	    NMF is a constrained dimensionality reduction technique which ensures nonnegativity hence is the natural choice for nonnegative network traffic flow estimation. It has already been successfully applied in many engineering applications, including but not limited to, computer vision, signal processing~\cite{leplat_blind_2020}, hyperspectral sensing~\cite{ang_algorithms_2019}, and background subtraction~\cite{chen_online_2016}. However to the best of our knowledge, we are the first to conduct a comprehensive research with conclusive results for the use of NMF in the context of network traffic flow estimation. Along with the nonnegativity constraint, we also impose two additional constraints on the proposed model for effectively capturing the spatial and temporal features in the traffic flows of the underlying network. 
    	The key  contribution of this paper is threefold: 
    	\begin{enumerate}[(1)]
    		\item A novel multi-constrained nonnegative matrix factorization model is proposed for network traffic flow estimation. The proposed model consists of a dimensionality reduction for capturing spatial features of the network traffic flows with a novel combination of three additional constraints:
    		\begin{inparaenum}[a)] 
    			\item nonnegativity,
    			\item autoregression, and
    			\item orthogonality. 
    		\end{inparaenum} 
    		The autoregression constraint allows a dynamic learning of autoregression weights and an effective representation of negative correlation between different OD flows which are prominent features of the proposed model and are not available in the conventional graph embedding approaches. 
    		The orthogonality constraint provides a distance preserving transformation from link flows to latent flows enabling effective clustering of the given link flows according to their temporal similarities.
    		
    		\item We propose an efficient  training algorithm using the fast gradient method of Nesterov, while we use a noise resilient initialization strategy that provides a deterministic seeding point to the training algorithm~\cite{atif_improved_2019}. 
    		
    		\item We illustrate the effectiveness of the proposed model and algorithm on two publicly available Internet backbone traffic data sets, namely  Internet2~\cite{yin_zhang_abilene_nodate} and G\'EANT~\cite{uhlig_providing_2006}. 
    		The empirical results show that the new model performs competitively with the current state-of-the-art models.  
    	\end{enumerate}
	
	    The rest of the paper is organized as follows. Section~\ref{sec:pro_sol} introduces the proposed model in detail, highlighting the key components and their advantages along with differences with existing approaches. Section~\ref{subsec:training_algo} presents our proposed algorithm to tackle our new model. In Section~\ref{sec:experimental_results}, the proposed model is extensively evaluated on two publicly available Internet traffic data sets, and we show that our proposed model competes favorably with state-of-the-art algorithms.  
	    
    	\section{New model for network traffic flow estimation}\label{sec:pro_sol}
    	
    	Let us formulate our proposed NMF-based network traffic estimation model for a network that is monitored for time period $T$, and consists of $m$ links and $n$ OD pairs. For this time period, we are given $X \in \mathbb{R}^{n \times T}$, and our goal is to construct a model allowing us to predict $x_t$ for $t > T$, given $A$ and $y_t$; see Section~\ref{sec_intro}. More precisely, we are given  \begin{itemize}
    	    \item the routing matrix $A \in \mathbb{R}^{m \times n}$, 
    	    
    	    \item the OD flow matrix $X \in \mathbb{R}^{n \times T}$, 
    	    
    	    \item a factorization rank $k$, and 
    	    
    	    \item  a lag set {$\mathcal{L}=\{l:l \in \mathbb{N} \text{ and } l \ll T\}$} 
    	with $L = \max(\mathcal{L})$. The lag set contains the indices $l$ indicating a dependency between the $t$th and $(t-l)$th time points, such as in autoregressive models; see Section~\ref{subsec:autoregression} for more details. 
    	\end{itemize}  
    	The model aims at computing $(W,H,\Omega)$ by solving the following optimization problem: 
    	\begin{subequations}\label{eq:proposed_model}
    		\begin{align}
    		\underset {W \in \mathbb R^{n \times k}, H \in \mathbb R^{k \times T}, \Omega \in \mathbb{R}^{k \times L}}{\min} 
    		 \| X-WH \| _F^2\label{subeq:datafitting}\\		
    		\text{such that } W \ge \boldsymbol{0}, H \ge \boldsymbol{0},\Omega \ge \boldsymbol{0}, \label{subeq:nonnegativity}\\		
    		H(p,t) = \sum_{l \in \mathcal{L}}\Omega(p,l)H(p,t-l)\nonumber \\
    		\text{ for } 1 \le p\le k, \; L < t \le T, \label{subeq:autoregression} \\ 
    		\Omega(p,l) > 0, \nonumber \\
    		\mathcal{A}= AW, \quad \mathcal{A}^{\top}\mathcal{A} = I_{k},\label{subeq:isometry}
    		\end{align}
    	\end{subequations}
    	where $\|.\|_{F}$ denotes the Frobenius norm, and the objective \eqref{subeq:datafitting} is the data fitting term of the model. 
    	Let us discuss the constraints in the above model: 
    	\begin{itemize}
    	    \item \eqref{subeq:nonnegativity} ensures nonnegative entries in $W$, $H$ and $\Omega$.  
    	    
    	    \item  \eqref{subeq:autoregression} imposes that each entry $H(p,t)$ of matrix $H$ is the nonnegative weighted sum of entries in the $p$th row of matrix $H$ preceding $H(p,t)$. This is motivated by the autoregression model; see Section~\ref{subsec:autoregression} for the details. 
    	    
    	    \item \eqref{subeq:isometry} defines a rank-$k$ compact routing matrix $\mathcal{A} = AW \in \mathbb{R}^{m \times k}$ that has orthonormal columns as $I_{k}$ stands for a $k \times k$ identity matrix; see Section~\ref{sec_ortho} for the details. 
    	    
    	\end{itemize}  
	
	    In the context of dimensionality reduction, the low-rank matrices $W$ and $H$ are generally called the basis and embedding (encoding) matrix, respectively. However, in the context of traffic matrix estimation, $W$ contains the spatial features of the OD flow matrix $X$ in low dimension, while $H$ can be interpreted as a latent flow matrix defining the latent flows of the underlying network. 

	    In order to incorporate the constraints \eqref{subeq:nonnegativity}-\eqref{subeq:isometry}, we will resort to regularization so that the constraints will not be strictly satisfied, but will  tend to be.
    	Using regularization, the optimization task in \eqref{eq:proposed_model} can be formulated as a regularized NMF problem; see Section~\ref{incorporation_of_constraints}.  
	
	    Let us now discuss the two key constraints in~\eqref{eq:proposed_model}, namely orthogonality~\eqref{subeq:isometry} and autoregression~\eqref{subeq:autoregression}. 
	
	        \subsection{Orthogonality constraint}  \label{sec_ortho}
	
	        The constraint $\mathcal{A}^T \mathcal{A} = I_k$  can be equivalently written as $W^T (A^T A) W = I_k$, which imposes $k^2$ constraints on $W \in \mathbb{R}^{n \times k}$. 
	        Note that $\mathcal{A} = AW$ is also nonnegative since $A$ and $W$ are, and hence this orthogonality constraint requires the columns of $\mathcal{A}$ to have disjoint supports (that is, the set of nonzero entries of the columns of $AW$ do not intersect). This is related to the so-called orthogonal NMF model; see~\cite{pompili2014two} and the references therein.   
	        Hence this constraint requires the support of the columns of $W$ to be disjoint as well, imposing $W$ to learn different features from the data set. Equivalently, it implies that there is at most a single non-zero entry in each row of $\mathcal{A}$ and $W$. 

        	Looking back at the model $Y = AX \approx (AW) H = \mathcal{A} H$, this means that each row of $Y$ is approximated as an scaling of a single row of $H$~\cite{pompili2014two}. 
        	Because we will not enforce~\eqref{subeq:isometry} strictly, but use a regularization, our minimization problem is related to soft clustering, which can be effectively used to capture diurnal similarity patterns (temporal dependencies) present in the link count matrix $Y$. 
	
        	\paragraph{Estimation of $\boldsymbol{x}_t$}  	 Another salient aspect of the proposed model is that the estimate of an OD flows, $\hat{\boldsymbol{x}}_{t}$ at time $t$, using the observed link count $\boldsymbol{y}_{t}$, is purely nonnegative. In fact, as explained in Section~\ref{sec_intro}, $\hat{\boldsymbol{x}}_{t}$ is estimated from the model  
        	\[ 
        	\boldsymbol{y}_t \approx A \boldsymbol{x}_t = A W \boldsymbol{h}_t = \mathcal{A} \boldsymbol{h}_t,
        	\] 
        	and by taking $\boldsymbol{h}_t = \mathcal{A}^\dagger \boldsymbol{y}_t$. We have 
        	$\mathcal{A}^\dagger = \mathcal{A}^T$ since 
        	$\mathcal{A}^T \mathcal{A} = I_k$, so that 
        	$\boldsymbol{x}_t = W\boldsymbol{h}_t = W \mathcal{A}^T \boldsymbol{y}_t$ which is nonnegative since $A, \mathcal{A}, \boldsymbol{y}_t \geq 0$.
        	This desirable feature of the proposed method is  due to the additional nonnegativity and orthogonality constraints,   \eqref{subeq:nonnegativity} and \eqref{subeq:isometry}. 
        	To the best of the authors' knowledge, this aspect has been missing in current state-of-the-art methods that generally employ an additional step of setting negative entries to zero. 

    	    \subsection{Temporal modeling using autoregression}\label{subsec:autoregression}
        	
        	The constraint~\eqref{subeq:autoregression} defines $k$ independent autoregression models for 
        	$k$ timeseries of latent flows corresponding to the $k$ rows of the matrix $H$. 
        	For the $p$th timeseries of latent flows, the autoregression model approximates every element in that timeseries as a weighted sum of its previous elements, that is, 
        	\[ 
        	H(p,t) \; = \;  \sum_{l \in \mathcal{L}}\Omega(p,l)H(p,t-l). 
        	\] 
        	When computing the weighted sum for an element $H(p,t)$, not all elements preceding it are taken into account but only the elements $H(p,t-l)$ for $l \in \mathcal{L}=\{l: l \in \mathbb{N} \text{ and } l \ll T\}$, where $\mathcal{L}$ is the lag set. 
        	In our model~\eqref{subeq:autoregression}, all $k$ autoregressive models share a common lag set $\mathcal{L}$, but there are $k$ weight vectors gathered in the matrix $\Omega$. The lag $\mathcal{L}$ have indices to indicate positive correlation among OD flows. In particular, the OD flows $\boldsymbol{x}_{t-l}$ and $\boldsymbol{x}_{t}$ are assumed to be correlated if $l \in \mathcal{L}$. Equivalently, this means that  the OD flow vectors $\boldsymbol{x}_{t-l}$ and $\boldsymbol{x}_{t}$ are assumed to be correlated, since our model assumes $\boldsymbol{x}_t \approx W\boldsymbol{h}_t$ for all $t$. 
    	
    	    Modeling temporal dependencies via~\eqref{subeq:autoregression}, that is, using $k$ autoregressive models sharing a common lag set $\mathcal{L}$, has several advantages over using a  multivariate autoregressive model or over conventional graph embedding approach: 
    	    \begin{enumerate}
    		    \item It requires only $k \times |\mathcal{L}|$ nonnegative weights in contrast to multivariate autoregressive models which require $k \times k \times |\mathcal{L}|$ nonnegative weights. Hence, it is less prone to overfitting and noise. 
    		
    		    \item There is no restriction on defining indices of the lag set $\mathcal{L}$. In contrast, conventional graph embedding approaches (e.g.\ \cite{deng_cai_graph_2011}) either use a lag set with few elements with short dependencies,  or require a prior knowledge of temporal dependencies which is generally not available.
    		
        		\item Unlike conventional graph embedding approaches (e.g.\ \cite{deng_cai_graph_2011}), weights of $k$ autoregression models (that is, $\Omega$) can be learnt dynamically.
        		
        		\item Like conventional graph embedding approaches such as~\cite{deng_cai_graph_2011}, each of the $k$ autoregression models have an equivalent graph representation as explained in~\cite{yu_temporal_2016}; see Figure~\ref{fig:temporal_regularizers}(a).
        		\begin{figure}[ht!]
        		\centering
        		    \begin{subfigure}[b]{0.48\textwidth}
        			    \centering         			    \includegraphics[width=\textwidth]{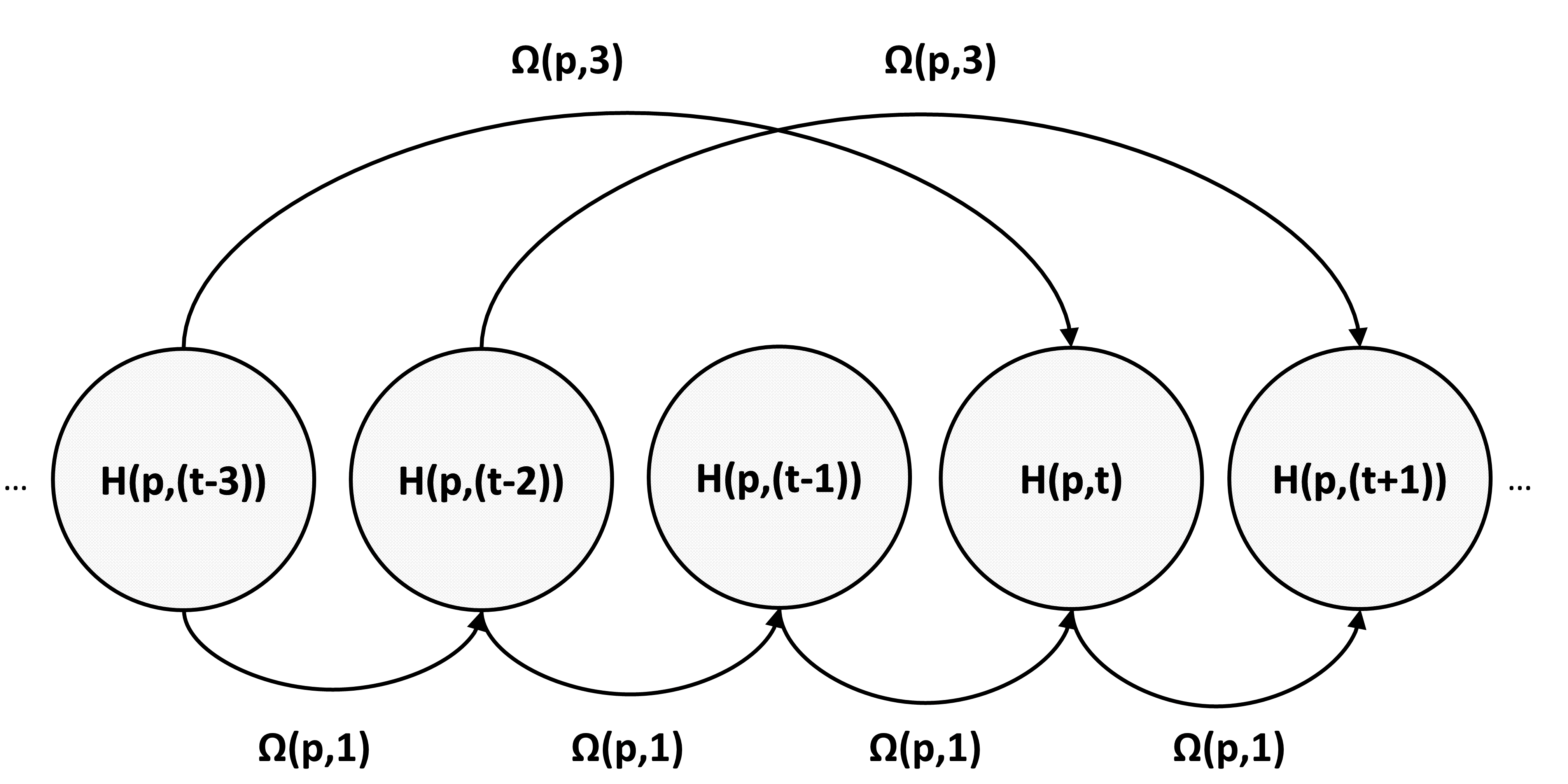}
        			    \caption{} \label{subfig:graph_based_temporal_ragualerizer}
        			\end{subfigure}
        			\hfill
        			\begin{subfigure}[b]{0.48\textwidth}
        			    \centering
        			    \includegraphics[width=\textwidth]{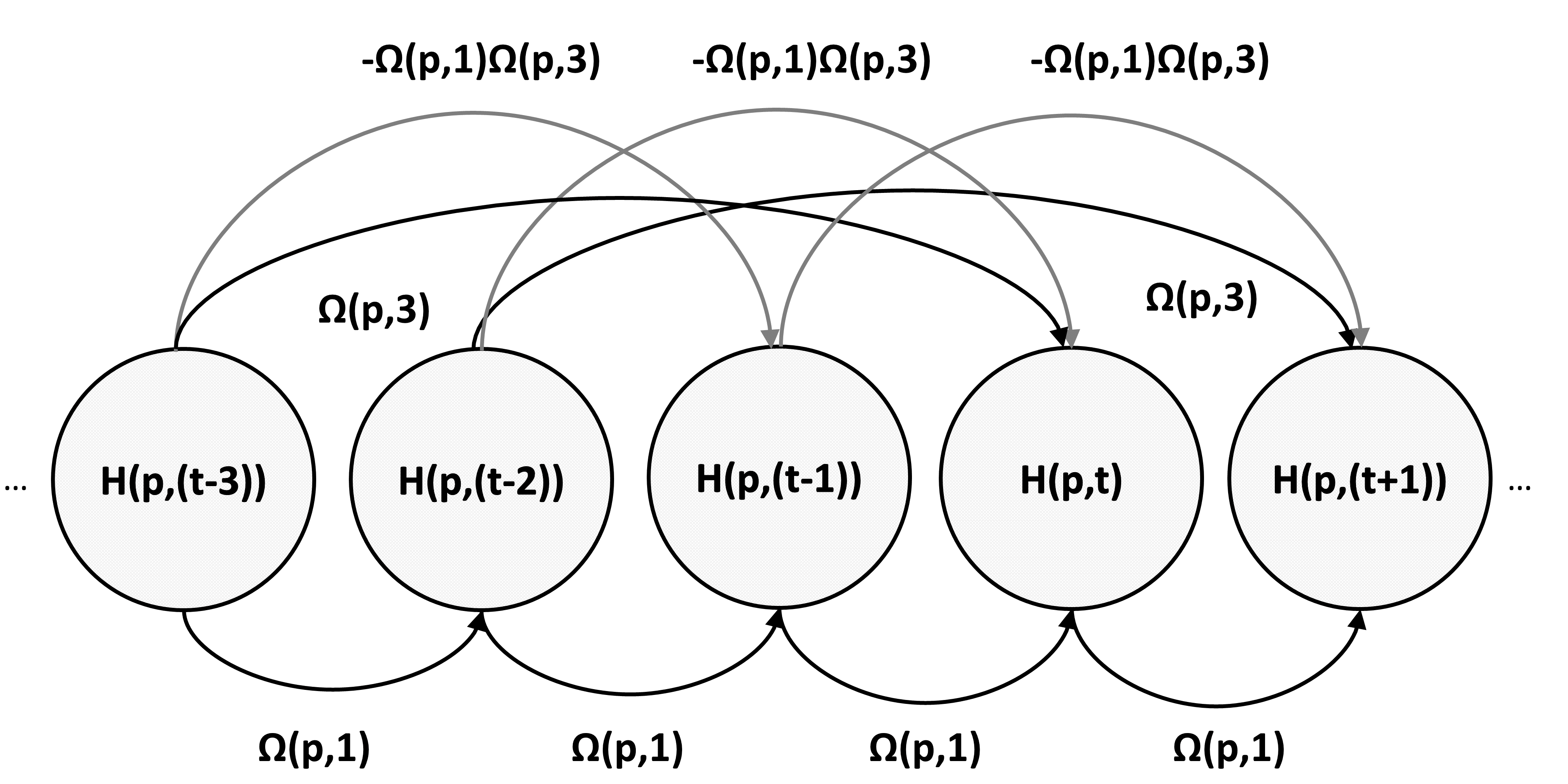}
        			    \caption{} \label{subfig:autoregression_temporal_ragualerizer}
        		    \end{subfigure}
        		\caption{
        			(a) A nonnegative weighted graph $\mathcal{G}$ defined by a typical graph embedding approach and associated to the $p$th row of $H$.
        			(b) A weighted signed graph $\mathcal{G}^{AR}$ defined by the $p$th autoregressive model $H(p,t) = \sum_{l \in \mathcal{L}}\Omega(p,l)H(p,t-l)$ associated to the $p$th row of $H$. In both cases the lag set is supposed to be $\mathcal{L}=\{1,3\}$. In (b) gray colored arrows indicates negatively weighted edges due to negative correlation between the two nodes. This figure is inspired
        			from~\cite[Figures 2-3]{yu_temporal_2016}. 
        		}\label{fig:temporal_regularizers}
        	\end{figure} 
        	Hence, the  constraint~\eqref{subeq:autoregression} can be incorporated into the data fitting term~\eqref{subeq:datafitting} using exiting techniques like Laplacian regularizer.
    		
    		    \item Unlike conventional graph embedding approaches such as~\cite{deng_cai_graph_2011}, the graph associated with the  $p$th autoregression model may contain negatively weighted edges to indicate negative correlation between two nodes of the graph; see  Figure~\ref{fig:temporal_regularizers}(b) for an illustration.     
    	    \end{enumerate}	
    	    We refer the interested readers to~\cite{yu_temporal_2016} and the references  for more details.  
		
	    \section{Training our traffic estimation model~\eqref{eq:proposed_model}} \label{subsec:training_algo}
		 
		 In this section, we first provide our regularized NMF model to tackle~\eqref{eq:proposed_model} in Section~\ref{incorporation_of_constraints}. In fact, in practice, because of noise and model misfit, it is not reasonable to strictly enforce the constraints~\eqref{subeq:autoregression}  and~\eqref{subeq:isometry} (in fact, the orthogonality constraint could even make the problem infeasible), and hence it makes more sense to only penalize the solutions that violate these constraints. 
		 Then we propose a fast gradient method to solve it in Section~\ref{section_fast_grad}. 
		 
		    \subsection{Incorporation of constraints into the model}\label{incorporation_of_constraints}
	
		    Let us consider the following regularized  NMF-based network traffic estimation model, which replaces the constraints~\eqref{subeq:autoregression}  and~\eqref{subeq:isometry} 
		    in~\eqref{eq:proposed_model} with  regularization terms:
            \begin{align}\label{eq:proposed_model_with_regularization}
				\underset {W \in \mathbb R^{n \times k}_+, H \in \mathbb R^{k \times T}_+, \Omega \in \mathbb{R}^{k \times L}_+}{\min} & F(W,H,\Omega), \\ 
				\text{ where } F(W,H,\Omega) = & \| X-WH \| _F^2 + \lambda_{h} \sum_{p=1}^{k}\mathcal{T}(\omega_{p},\mathcal{L},h_{p}) + \lambda_{\mathcal{A}} \mathcal{I}(\mathcal{A}),\nonumber 
		    \end{align}
            where  $\mathcal{T}(\omega_{p},\mathcal{L},h_{p})$ is the temporal regularizer to incorporate the constraint~\eqref{subeq:autoregression} with  $\omega_{p}$ (resp.\ $h_{p}$) the $p$th row of  $\Omega$ (resp.\ $H$), and  $\mathcal{I}(\mathcal{A})$ is the orthogonality regularizer to incorporate the constraint~\eqref{subeq:isometry}; see below for more details. 
            The positive numbers $\lambda_{h}$ and $\lambda_{\mathcal{A}}$ are the penalty parameters for the regularizers $\mathcal{T}(\omega_{p},\mathcal{L},h_{p})$ and $\mathcal{I}(\mathcal{A})$ respectively and, $\omega_{p}$ and $h_{p}$ are the $p$th rows of $\Omega$ and $H$ respectively. 
 
		    Let us briefly discuss the two regularizers. 
		
		    \paragraph{Orthogonality} 
		    The orthogonality regularizer $\mathcal{I}(\mathcal{A})$ in \eqref{eq:proposed_model_with_regularization} corresponds to the {orthogonality} constraint~\eqref{subeq:isometry} and is defined as:
			\[
			    \mathcal{I}(\mathcal{A}) = \|\mathcal{A}^{\top}\mathcal{A} - I\|_{F}^{2}. 
			\] 
			This is standard least-squares penalty for orthogonality constraints; see for example~\cite{ahookhosh2019multi}. 
			
			\paragraph{Autoregression} 
		    We use the $p$th temporal regularizer $\mathcal{T}(\omega_{p},\mathcal{L},h_{p})$ from \cite{yu_temporal_2016}, which is   equivalent to the following Laplacian regularizer: 
    		\begin{multline}\label{eq:autoregressive_temporal_regularizer}
    		\mathcal{T}(\omega_{p},\mathcal{L},h_{p}) = \frac{1}{2}\sum_{t_{1},t_{2}}^{T}S_{p}^{AR}(t_{1},t_{2})\|H(p,t_{1})-H(p,t_{2})\|^{2} + \frac{1}{2}\boldsymbol{h}_{p}D_{p}\boldsymbol{h}_{p}^{\top}
    		\end{multline} 
    		where $S_{p}^{AR} \in \mathbb{R}^{T \times T}$ is the weight matrix of the graph $\mathcal{G}^{AR}$ and $D_{p} \in \mathbb{R}^{T \times T}$ is the diagonal matrix such that:
    		\[
    		S_{p}^{AR}(t,t+d) = 
    		\begin{cases}
    		\sum_{l \in \delta(d)}\sum_{L < t \le T}-\Omega(p,l)\Omega(p,(l-d)),  &\text{if } \delta(d) \neq \phi,\\
    		0, &\text{otherwise,}
    		\end{cases}
    		\]
    		where $\delta(d)=\{l \in \mathcal{L} \cup \{0\} : l-d \in  \mathcal{L} \cup \{0\}\}$ and $L = \max(\mathcal{L})$, 
    		and 
    		\[
    		D_{p}(t,t) = \Bigg(\sum_{l \in \mathcal{L} \cup \{0\}}\Omega(p,l)\Bigg)\Bigg(\sum_{l \in \mathcal{L} \cup \{0\}}\Omega(p,l)[L < t+l \le T]\Bigg) . 
    		\]
    		The graph Laplacian $L_{p}^{AR} \in \mathbb{R}^{T \times T}$ is defined as:
    		\[
    		    L_{p}^{AR}(t_{1}, t_{2}) = 
    		    \begin{cases}
    		           \sum_{t_{3} = 1}^{T} S_{p}^{AR}(t_{1}, t_{3}) & \text{ if } t_{1} = t_{2},\\
    		           -S_{p}^{AR}(t_{1}, t_{2}) & \text{ otherwise.}
    		    \end{cases}
    		\]
    		The first term in \eqref{eq:autoregressive_temporal_regularizer} is standard for Laplacian regularizers of conventional graph embedding approaches (see, e.g., \cite{deng_cai_graph_2011}), 
    		whereas the second term is specific to the  autoregression model~\cite{yu_temporal_2016} due to negative edges in the associated graph $\mathcal{G}^{AR}$ indicating negative correlation between nodes; 
    		see~ Figure~\ref{subfig:autoregression_temporal_ragualerizer}.

            \subsection{Fast gradient algorithm} \label{section_fast_grad}
        		
    		In this section, we describe a training algorithm for the network traffic estimation model \eqref{eq:proposed_model_with_regularization}. The algorithm is iterative and is based on block coordinate descent method, as most NMF algorithms~\cite{gillis2020book}.  
    		Such iterative algorithms require an initialization. 
    		The initial iterates are denoted $W^{(0)}$, $H^{(0)}$ and $\Omega^{(0)}$
    		which we compute as follows: 
    		\begin{enumerate}\label{enum:init_strategy}
    		
    			\item For the given training OD flow matrix $X$ and the rank of factorization $k$, the matrices $W$ and $H$ are initialized by ($W^{(0)}$, $H^{(0)}$) = NNSVD-LRC($X$, $k$) where NNSVD-LRC is an effective initialization for NMF based on the SVD~\cite{atif_improved_2019}.
    			
    			\item The $p$th row $\omega_{p}$ of $\Omega$ is initialized as the projection of the unconstrained solution: 
    			\[ \omega^{(0)}_{p} = \mathcal{P}_{+}(h_{p} \mathcal{H}_{p}^{\dagger}) , 
    			\] 
    			where $\mathcal{P}_{+}(Z)$ is the projection onto the nonnegative orthant, that is, 
    			\[
    			\mathcal{P}_{+}(Z)_{i,j} = 
    			\begin{cases}
    			Z(i,j),  & \text{if } Z(i,j) \ge 0, \\
    			0, &  \text{otherwise}, 
    			\end{cases} 
    			\]
    		where $\mathcal{P}_{+}(Z)_{i,j}$ denotes the entry of $\mathcal{P}_{+}(Z)$ at position $(i,j)$, $\mathcal{H}_{p}^{\dagger}$ is the (left) pseudoinverse of $\mathcal{H}_{p}$  
    			  and $\mathcal{H}_{p} \in \mathbb{R}^{L \times T}$ is defined as: 
    			\[
    			\mathcal{H}_{p}(q,t)=
    			\begin{cases}
    			H(p,(t-q)), & \text{if }q \in \mathcal{L} \text{ and } L < t \le T\\
    			0, & \text{otherwise.} 
    			\end{cases}
    			\]	
    		\end{enumerate} 

            \begin{algorithm}[ht!]
    			\caption{Solving the traffic matrix estimation model~\eqref{eq:proposed_model_with_regularization}, \textsc{MCST-NMF}}
    			\label{algo:mcst_nmf_training}
                \begin{algorithmic}[1]
    				\REQUIRE 
    				$X$ is an $n$-by-$T$ training OD flow matrix, 
    				$W^{(0)}$,$H^{(0)}$ and $\Omega^{(0)}$ are the initial values for $W$,$H$ and $\Omega$ respectively, 
    				$q^{max}$ is the maximum number of iterations, 
    				and $0 \leq \delta \ll 1$ is a threshold for early stopping when the differences between two consecutive errors is small enough. 
    				\ENSURE Final values for $W$, $H$ and $\Omega$ and $\epsilon^{min}$ solving~\eqref{eq:proposed_model_with_regularization}	 	
    				
    				\STATE	Initialize: $W = W^{(0)}$; $H = H^{(0)}$; $\Omega = \Omega^{(0)}$; $e(0) = \|X-WH\|_{F}^{2}$;
    				
    				$\epsilon_{min} = \delta e(0) $; $\epsilon = \epsilon^{min}$; $q = 1$;
    				\REPEAT
    				    \STATE $W =$ FastGradientUpdate($X, W, H, \Omega$);
    					\STATE $H =$ FastGradientUpdate($X, W, H, \Omega$);
    					\STATE $\Omega =$ FastGradientUpdate($X, W, H, \Omega$);
    										
    					\STATE $e(q) = \|X-WH\|_{F}^{2}$;
    					$\epsilon = e(q-1) - e(q)$; $q=q+1$;
    				\UNTIL $q \le q^{max} \text{ and } \big( \epsilon < 0 \text{ or } \epsilon \ge \epsilon^{min} \big)$ 
    				
    			\end{algorithmic}
    		\end{algorithm}
		    		
		    Given the initial estimate $(W^{(0)}, H^{(0)}, \Omega^{(0)})$, Algorithm~\ref{algo:mcst_nmf_training} further improves the solution via an iterative process.  We use a standard strategy in NMF, that is, alternatively updating the block variables $(W,H,\Omega)$. For each block, we use a first-order accelerated gradient descent method with optimal convergence rate~\cite{nesterov_method_nodate}; as done in~\cite{guan_nenmf:_2012} for the standard NMF model. 
		    In a nutshell, such methods take a gradient step from an extrapolated sequence.
		    
    	    Algorithm~\ref{algo:mcst_nmf_training} describes the proposed training algorithm in detail. It takes as as input the initial values of three free parameters $W^{(0)}$, $H^{(0)}$ and $\Omega^{(0)}$, the training OD flows $X$, the routing matrix $A$, the rank of factorization $k$, the lag set $\mathcal{L}$, and the maximum number of iterations $n_{max}$. In each iteration Algorithm~\ref{algo:mcst_nmf_training} alternatively optimizes $W$, $H$ and $\Omega$ using the accelerated gradient method as described in Algorithm~\ref{algo:single_update_WHOmega} with help of Table~\ref{tb:fast_gradient_update}. Note that we use a restarting mechanism in the fast gradient method which typically performs well in practice; see the discussion in~\cite{o2015adaptive}.   
    	    
    	   \begin{table}
			\caption{Computations needed for Algorithm~\ref{algo:single_update_WHOmega}, FastGradientUpdate($X, W, H, \Omega$), depending on the updated variable, $B$ is either $W$, $H$ or $\Omega$.} 
			\label{tb:fast_gradient_update}
			\resizebox{\columnwidth}{!}{
			\begin{tabular}{l l l l}
				\toprule[1.5pt]
				variable & quantity & formula \\
				\midrule[1.5pt]
				& $L_{B}$ & $\| W^{\top}W \|_{2}$ \\
				& $\nabla_{B}F(W,H,\Omega)$ & $2(WHH^{\top} - XH^{\top}) + 4\lambda_{\mathcal{A}}(A^{\top}\mathcal{A})(\mathcal{A}^{\top}\mathcal{A} - I_{k})$ \\
				\multirow{-3}{*}{$W$}& $e_{B}^{(curr)}$ & $\|X-WH\|_{F}^{2}$\\
				\midrule[0.75pt]
				& $L_{B}$ & $\| HH^{\top} \|_{2} + \sum_{p = 1}^{k} \lambda_{h} ( \| L_{p}^{AR} \|_{2} + \| D_{p} \|_{2} )$ \\
				& $\nabla_{B}F(W,H,\Omega)$ & $2(W^{\top}WH - W^{\top}X) + \lambda_{h}H(p,:)\Big((L_{p}^{AR})^{\top} + L_{p}^{AR}\Big)$ \\
				\multirow{-3}{*}{$H$}& $e_{B}^{(curr)}$ & $\|X-WH\|_{F}^{2}$\\
				\midrule[0.75pt]
				& $L_{B(p,:)}$ & $\| \mathcal{H}_{p}\mathcal{H}_{p}^{\top} \|_{2}$, \hfill for $p = 1, 2, \dots, k$.\\ 
				& $\nabla_{B(p,:)}F(W,H,\Omega)$ & $2\Big(\Omega(p,:)\mathcal{H}_{p}\mathcal{H}_{p}^{\top} - H(p,:)\mathcal{H}_{p}^{\top}\Big)$ \\
				\multirow{-3}{*}{$\Omega$} & $e_{B}^{(curr)}$ & $\sum_{p=1}^{k}\|H(p,:) - \Omega(p,:)\mathcal{H}_{p}\|_{2}^{2}$\\
				\bottomrule[1.5pt]
			\end{tabular}
			}
			\end{table}

    	   \begin{algorithm}[ht!]
    			\caption{FastGradientUpdate($X, W, H, \Omega$)} \label{algo:single_update_WHOmega}
               \begin{algorithmic}[1] 
    				\REQUIRE $X$ is an $n$-by-$T$ training OD flow matrix, $W$, $H$, and $\Omega$ are the three free parameters, $q_{B}^{max}$ the maximum iterations for loop, and $1 \gg \delta_{B} > 0$ the minimum differences between two consecutive errors w.r.t.\ initial error for early stopping before $q_{B}^{max}$ iterations respectively.
    				\ENSURE Updated $B$ by solving~\eqref{eq:proposed_model_with_regularization} for $B$ (where $B$ can be $W$, $H$ or $\Omega$)	
    				\STATE Initialize: $C=B$; $B^{(curr)}=B$; $\alpha_{B} = \alpha_{B}^{(prev)} = 1$; $q_{B} = 1$; \STATE Compute $e_{B}^{(curr)}$ as described in Table~\ref{tb:fast_gradient_update}
    				\STATE $e_{B}^{(prev)} = e_{B}^{(curr)}; \epsilon_{B}^{min} = \delta_{B}e_{B}^{(prev)}$;
    				\REPEAT
    				    \STATE $\alpha_{B} = \frac{1 + \sqrt{4\alpha_{B}^{2} + 1}}{2}$;
    				    \STATE Compute $L_{B}$ and $\nabla_{B}F(W,H,\Omega)$ as described in Table~\ref{tb:fast_gradient_update}.
    				    \STATE $B = \mathcal{P}_{+}\Big(C - \frac{1}{L_{B}}\nabla_{B}F(W,H,\Omega)\Big)$;
        				\STATE $C = B + \frac{\alpha_{B}^{(prev)} - 1}{\alpha_{B}}\Big(B - B^{(curr)}\Big)$;
        				\STATE Compute $e_{B}^{(curr)}$ as described in Table~\ref{tb:fast_gradient_update}
    					\STATE $\epsilon_{B} = e_{B}^{(prev)} - e_{B}^{(curr)}$;
    					\IF{$\epsilon_{B} < 0$}
    					\STATE $C = B$; $\alpha_{B} = 1;$ $e_{B}^{(curr)} = e_{B}^{(prev)}$;
    					\ENDIF
    				
    					\STATE $\alpha_{B}^{(prev)} = \alpha_{B}$; $e_{B}^{(prev)} = e_{B}^{(curr)}$; $q_{B}=q_{B}+1$; 
    				\UNTIL $q_{B} \le q_{B}^{max} \text{ and } \big( \epsilon_{B} < 0 \text{ or } \epsilon_{B} \ge \epsilon_{B}^{min} \big)$
    					
    			\end{algorithmic}
    		\end{algorithm}

    	    Because accelerated gradient methods do not ensure the objective function to decrease at every iteration, and because we optimize alternatively $(W,H,\Omega)$, we embed Algorithm~\ref{algo:single_update_WHOmega} with a restarting scheme: if the objective function increases, the algorithm abandon the extrapolated sequence and takes a standard gradient step, as done for example in~\cite{xu2013block}. 
    		This ensures the objective function will decrease.
    		
    		
    		
    
            \paragraph{Choice of the penalty parameters}\label{para:penalty_parameter_choice} 
    		It is not an easy task to tune the two penalty parameters $\lambda_{h}$ and $\lambda_{\mathcal{A}}$ for the regularization terms in \eqref{eq:proposed_model_with_regularization}. 
    		Given the initial iterate $(W^{(0)},H^{(0)},\Omega^{(0)})$, we will use 		
    		\begin{subequations}\label{eq:tune_lambdas}
    			\begin{align}
    				\lambda_{h} & = \beta_{h} \frac{ \|X - W^{(0)}H^{(0)}\|_{F}^{2} }{ \sum_{p=1}^{k} \| h_{p} - \omega_{p}^{(0)}\mathcal{H}_{p}^{(0)} \|_{2}^{2} }\label{subeq:tune_lambda_h}, \\
    				\lambda_{\mathcal{A}} & = \beta_{\mathcal{A}} \frac{ \|X - W^{(0)}H^{(0)}\|_{F}^{2} }{ \| \big(\mathcal{A}^{(0)}\big)^{\top}\mathcal{A}^{(0)} - I \|_{2}^{2}}, \label{subeq:tune_lambda_A}
    			\end{align}
    		\end{subequations}
    		where $\beta_{h}, \beta_{\mathcal{A}} \in (0,1]$. 
    		This choice allows to balance the importance of the penalty terms compared to the data fitting term, at  initialization. Like other existing regularized models, e.g.,~\cite{roughan_spatio-temporal_2012}, 
    		the tuning of the penalty parameters, $\beta_{h}$ and $\beta_{\mathcal{A}}$  in our model, is a difficult task and typically problem dependent (nature of the given network, noise level, etc.). In practice, a useful way to tune  parameters is to use cross validation (see, e.g.,~\cite{crossvalid}), that is, use training sets to train a model with different parameter values, and then select the values of the parameters that lead to the best results on the test sets. 
    		For our problem, a value of 
    		$\beta_{h}$ and $\beta_{\mathcal{A}}$ around 10-20\% typically works well in practice. 
    		
    		\paragraph{Handling the missing entries} 
    		It is usual that the traffic matrix $X$ used for training a given traffic estimation model contains  missing entries, even in the presence of a good measurement system~\cite{tune_internet_2013}. To the best of our knowledge, all the existing state-of-the-art algorithms assume that the training data is full and complete. To address the issue of an incomplete training dataset, we propose two approaches. The first approach preprocesses the given training dataset to fill in the missing entries using a weighted NMF model trained using a fast gradient method~\cite{dorffer2017fast}. 
    		The second approach modifies  Algorithm~\ref{algo:mcst_nmf_training} using the expectation maximization strategy proposed in~\cite{zhang2006learning}. 
    		
    		Denoting $M \in \{0,1\}^{n \times T}$ the binary mask matrix so that $M(i,j) = 1$ if and only if $X(i,j)$ is observed, the second approach for handling the missing entries modifies Algorithm~\ref{algo:mcst_nmf_training} to use the data matrix $X^{(q)}$ instead of $X$ at iteration $q$,  where $X^{(q)}$ is defined as  
    		\[
    			X^{(q)} =  
    			    M \circ X + (\mathbbm{1}_{n \times T} - M) \circ (W^{(q-1)}H^{(q-1)})
    		\]
    		where $\circ$ is the element-wise matrix multiplication and $\mathbbm{1}_{n \times T}$ is the $n \times T$ all one matrix.  
    

     	\subsection{Estimation of future traffic flows}
     	 	  
    	\begin{algorithm}[ht!]
    		\caption{Estimating OD flows}
    		\label{algo:mcst_nmf_estimation} 
    		\begin{algorithmic}[1]
    				\REQUIRE $A$ an $m$-by-$n$ routing matrix, $W$ the $n$-by-$k$ matrix, the new observed link flow $\boldsymbol{y}_{t}$, $q^{max}$, and $r^{max}$ the maximum number of iterations  for fast gradient descent and expectation maximization iteration steps respectively, $1 \gg \delta_{gd}, \delta_{emi} > 0$ are the minimum difference between two consecutive errors for early stopping before $q^{max}$ and $r^{max}$ iterations of fast gradient descent expectation maximization iteration steps, respectively. 
    				
    				\ENSURE Final estimated OD flow $\hat{\boldsymbol{x}}_{t}$ for timestamp $t$.
    				
    				\STATE	Initialization for fast gradient steps: \\
    				$\mathcal{A} = AW$;
    				$\boldsymbol{h}_{t} = \mathcal{A}^{\top} \boldsymbol{y}_{t}$; $\boldsymbol{v}_{t} = \boldsymbol{h}_{t}$; $\boldsymbol{h}_{t}^{(curr)} = \boldsymbol{h}_{t}$;\\ $\alpha_{gd}^{(prev)} = 1$; 				
    				$\eta = \frac{ 1 }{ \| \mathcal{A}^{\top} \mathcal{A} \|_{2} }$; $e_{gd}^{(prev)} = \| \boldsymbol{y}_{t} - A \boldsymbol{h}_{t} \|_{2}^{2}$;\\
    				$\epsilon_{gd}^{min} = \delta_{gd} \|y_{t}\|_{2}^{2} $; $q = 1$;
    				\REPEAT
    					\STATE $\alpha_{gd} = \frac{1 + \sqrt{4\alpha_{gd}^{2} + 1}}{2}$;
    					\STATE $\boldsymbol{h}_{t} = \mathcal{P}_{+} \Big( \boldsymbol{v}_{t} - \eta \nabla_{\boldsymbol{h}_{t}} G(\boldsymbol{y}_{t}, \mathcal{A}, \boldsymbol{v}_{t} ) \Big)$;
    					\STATE $\boldsymbol{v}_{t} = \boldsymbol{h}_{t} + \frac{\alpha_{gd}^{(prev)} - 1}{\alpha_{gd}}\Big(\boldsymbol{h}_{t} - \boldsymbol{h}_{t}^{(curr)}\Big)$;
    					\STATE $e_{gd}^{(curr)} = \| \boldsymbol{y}_{t} - A \boldsymbol{v}_{t} \|_{2}^{2}$;
    					$\epsilon_{gd} = e_{gd}^{(prev)} - e_{gd}^{(curr)}$;
    					\IF{$\epsilon_{gd} < 0$} 
    					\STATE $\boldsymbol{v}_{t} = \boldsymbol{h}_{t}$; $\alpha_{gd} = 1$; $e_{gd}^{(curr)} = e_{gd}^{(prev)}$;
    					\ENDIF
    					\STATE $\alpha_{gd}^{(prev)} = \alpha_{gd}$; $e_{gd}^{(prev)} = e_{gd}^{(curr)}$; $q=q+1$;
    				
    				\UNTIL $q \le q^{max} \text{ and } \big( \epsilon_{gd} < 0 \text{ or } \epsilon_{gd} \ge \epsilon_{gd}^{min} \big)$ 
    
    				\STATE	Initialization for expectation maximization steps: \\
    				$\boldsymbol{x}^{(curr)} = W \boldsymbol{h}_{t}$; 
    				$\boldsymbol{y} = \boldsymbol{y}_{t}$; \\				
    				$e_{emi}(0) = 0$;
    				$\epsilon = \epsilon_{min}$; $\epsilon_{emi}^{min} = \delta_{emi} \|\boldsymbol{x}^{(curr)}\|_{2}^{2}$; $r = 1$;
    				
    				\REPEAT
    					\FOR{$j=1:1:n$}
    						\STATE $\boldsymbol{x}(j) = \frac{ \boldsymbol{x}(j)^{(curr)} }{ \sum_{i=1}^{m}A(i,j) } \sum_{i=1}^{m} \frac{ A(i,j) {\boldsymbol{y}(i)} }{ \sum_{k=1}^{n}A(i,k) \boldsymbol{x}(k)^{(curr)} }$
    					\ENDFOR
    				\STATE $\epsilon_{emi} = \| \boldsymbol{x} - \boldsymbol{x}^{(curr)} \|_{2}^{2}$; $\boldsymbol{x}^{(curr)} = \boldsymbol{x}$; $r=r+1$; 
    				
    				\UNTIL $r \le r^{max} \text{ and } (\epsilon_{emi} \ge \epsilon_{emi}^{min})$
    				
    				\STATE $\hat{\boldsymbol{x}}_{t} = \boldsymbol{x}$; 
    		\end{algorithmic}
    	\end{algorithm}
     	
     	It is a common practice to improve the solution obtained by the traffic estimation model using an Expectation Maximization (EM) iterative algorithm or iterative proportional fitting (IPF) algorithm; see for example \cite{soule_traffic_2005,jiang_accurate_2011,zhou_traffic_2016}. In this paper, we use the EM approach from \cite{vardi_1996_network} combined with a fast gradient method, which is empirically found to be more effective than applying the EM algorithm alone.  Algorithm~\ref{algo:mcst_nmf_estimation} provides the exact details of our traffic estimation procedure.
     	
     	Once the proposed model \eqref{eq:proposed_model_with_regularization} has been trained using Algorithm~\ref{algo:mcst_nmf_training} over a given network, let us explain how the new unobserved OD flows $\boldsymbol{x}_{t}$ can be estimated as $\hat{\boldsymbol{x}}_{t}$ using the estimated parameters $W$, the routing matrix $A$, and the observed link flow $\boldsymbol{y}_{t}$; see Algorithm~\ref{algo:mcst_nmf_estimation}.  
     	The algorithm first estimates the latent flow $h_{t}^{(0)}$ as $\mathcal{A}^{T}\boldsymbol{y}_{t}$; see Section~\ref{sec_ortho}. It is then refined using a few steps of projected fast gradient descent applied on the following minimization problem:
     	\begin{equation*}
     		\min_{ \boldsymbol{h}_{t} \in \mathbb{R}_{+}^{k} } G( \boldsymbol{y}_{t} , \mathcal{A} , \boldsymbol{h}_{t} ) = \min_{ \boldsymbol{h}_{t} \in \mathbb{R}_{+}^{k} } \| \boldsymbol{y}_{t} - \mathcal{A}\boldsymbol{h}_{t} \|_{2}^{2}. 
     	\end{equation*} 
    	Using the final estimated latent flow $\boldsymbol{h}_{t}^{(q_{f})}$ after $q_{f}$th iteration of the projected fast gradient method, initialize $\boldsymbol{x}^{(0)} = W \boldsymbol{h}_{t}^{(q_{f})}$ and $\boldsymbol{y} = \boldsymbol{y}_{t}$. The $r$th iteration of EM  for solving \eqref{eq:traffic_tomography_column} is as follows~\cite{vardi_1996_network,vardi_1996_applications}:
    	\begin{align*}
    	\boldsymbol{x}(j)^{(r)} = \frac{ \boldsymbol{x}(j)^{(r-1)} }{ \sum_{i=1}^{m}A(i,j) } \sum_{i=1}^{m} \frac{ A(i,j) \boldsymbol{y}(i) }{ \sum_{k=1}^{n}A(i,k) \boldsymbol{x}(k)^{(r-1)} },\quad \forall~j = 1,2, \dots ,n, 
    	\end{align*}
    	where $\boldsymbol{x}(j)^{(r-1)}$ and $\boldsymbol{x}(k)^{(r-1)}$ are the $j$th and $k$th OD flows of $\boldsymbol{x}^{(r-1)}$ respectively, $\boldsymbol{y}(i)$ is the $i$th link flow of $\boldsymbol{y}$ and $A(i,j)$ is the $(i,j)$th element of the binary routing matrix $A$. The output of the EM algorithm after $r_{f}$th iteration will be the final estimated OD flow  $\hat{\boldsymbol{x}}_{t} = \boldsymbol{x}^{(r_{f})}$ and be the output of Algorithm~\ref{algo:mcst_nmf_estimation}.
 	
	\section{Experimental Results}\label{sec:experimental_results}
	In this section, we evaluate the performance of our proposed traffic estimation model which we refer to as MCST-NMF (if the training dataset has missing entries, we use W-NeNMF~\cite{dorffer2017fast} by default to fill in the missing entries), otherwise as MCST-NMC when missing entries are dealt with using EM.  
	Moreover, we consider the following three state-of-the-art methods for the performance comparison 
	\begin{inparaenum}[a)]
	     \item PCA~\cite{soule_traffic_2005},
	     \item MNETME~\cite{zhou_traffic_2016}, and
	     \item CS-DME~\cite{qazi_novel_2018}.
	\end{inparaenum}
	We consider these state-of-the-art because they are all based on linear dimensionality reduction, like MCST-NMF; see Section~\ref{sec_intro}.  
	  
	\paragraph{Performance metrics}
	Given the true $X$ the true OD flow matrix of a test data and the corresponding estimated OD flow matrix $\widehat{X}$ by an algorithm, we use the two performance metrics SRE and TRE. They are row and column vectors of dimension $n$ and $T$ respectively. The $i$th and $t$th elements of SRE and TRE vectors are defined as follows:
	
	\begin{equation} \label{eq:SRE}
	SRE(i)=\frac{\sqrt{\sum_{t=1}^{T}\Big(\widehat{X}_{i,t}-X_{i,t}\Big)^{2}}}{\sqrt{\sum_{t=1}^{T}X_{i,t}^{2}}}, 
	\end{equation} 
	\begin{equation} \label{eq:TRE} 
	TRE(t)=\frac{\sqrt{\sum_{i=1}^{n}\Big(\widehat{X}_{i,t}-X_{i,t}\Big)^{2}}}{\sqrt{\sum_{i=1}^{n}X_{i,t}^{2}}}. 
	\end{equation}
	
	The scalar SRE$(i)$ is the normalized mean squared error of $i$th OD flow among all $T$ test timestamps, whereas TRE$(t)$ is the normalized mean squared error of the $t$th test timestamp among all $n$ OD flows.
	
	These two performance metrics, SRE and TRE, are widely used in the literature to measure the performance a given network traffic estimation model. SRE$(i)$ primarily describes how well a given model is able to estimate a given OD flow $i$ over a given time period $T$. Hence, it provides us the insight about the performance of the given model w.r.t.\ a given OD flow. In contrast, TRE$(t)$ describes the efficiency of a given model to estimate all OD flows in the given network at a particular timestamp $t$. Hence, it provides us with the insight about the performance of the given model w.r.t.\ a given timestamp. 
	
	Tables~\ref{tb:abilene_sre_tre} and~\ref{tb:geant_sre_tre} report the statistical properties of these two vectors, that is, of the SRE and TRE on the test dataset.

	\paragraph{Datasets}
	
\begin{figure*}[ht!]
		\centering
		\begin{subfigure}[b]{0.45\textwidth}
	        \centering
		    \includegraphics[width=\textwidth]{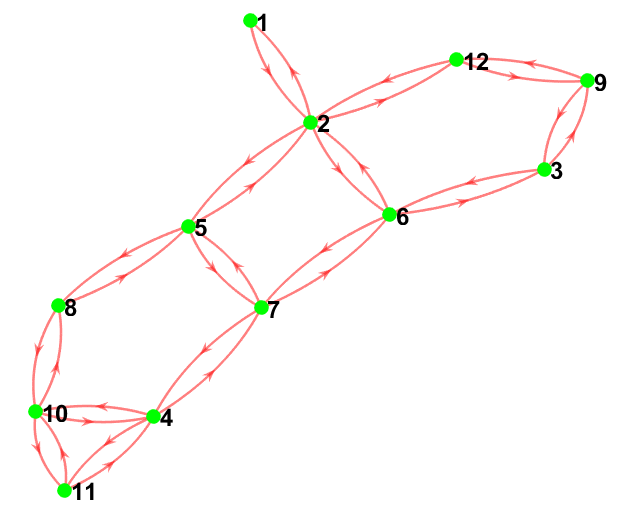}
		    \caption{Internet2 topology}
			\label{subfig:abilene_topology}
		\end{subfigure}
		\hfill
		\begin{subfigure}[b]{0.45\textwidth}
		    \centering
		    \includegraphics[width=\textwidth]{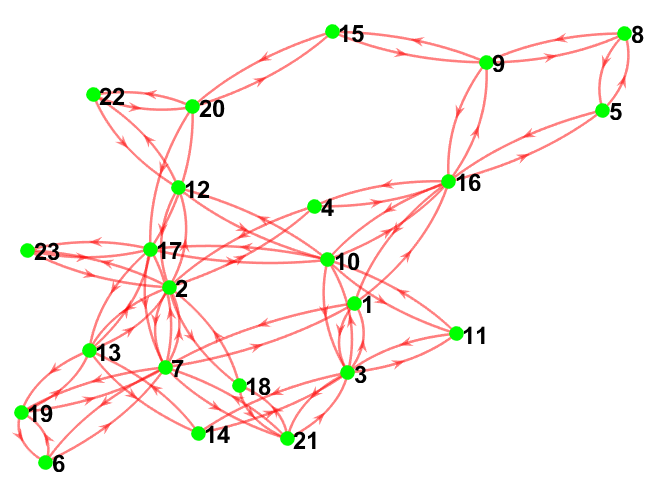}
		    \caption{G\'EANT topology}
		    \label{subfig:geant_topology}
		\end{subfigure}
		\caption{(a) Internet2 topology. (b) G\'EANT topology. Routers are numbered green coloured circles. Links are red coloured arrows. Only internal links are shown.}\label{fig:abilene_geant_topologies}
	\end{figure*}

	The performance of different models is evaluated over two publicly available and widely used Internet traffic datasets, namely  Internet2~\cite{yin_zhang_abilene_nodate} and G\'EANT~\cite{uhlig_providing_2006}. Internet2 (a.k.a.\ Abilene) is the high-speed backbone network of US. It consists of $12$ routers, $144$ OD flows and, $15$ and $12$ bidirectional internal and external links respectively. The topology of Internet2 network is depicted in Figure~\ref{fig:abilene_geant_topologies}-(a). The dataset~\cite{yin_zhang_abilene_nodate} provides the measurements of $24$ weeks of OD flows and the routing matrix of the network.  The traffic is measured at an interval of $5$ min and reported in the unit of $100$ bytes. G\'EANT is a research and educational network for Europe which is almost twice as large as Internet2. Figure~\ref{fig:abilene_geant_topologies}-(b) illustrates the topology of the network. It has $23$ routers, $23 \times 23 = 529$ OD flows and, $37$ and $23$ bidirectional internal and external links respectively. G\'EANT~\cite{uhlig_providing_2006} is a  collection of 4 months of OD flows measurements and the routing information of the network. The corresponding link flow matrix $Y$ can be obtained using~\eqref{eq:traffic_tomography_matrix}. Traffic measurements are in unit of kpbs and were performed 
	at an interval of $15$ minutes. Note that, in both cases, the link flow matrix $Y$ can be calculated using~\eqref{eq:traffic_tomography_matrix}.
	
	\paragraph{Parameter setting for experiments}\label{para:parameter_setting_for_experiments} 
		For experiments on the two datasets Internet2 and G\'EANT, we use the following specifications: 
		\begin{enumerate}[i)]
			\item For Internet2, the initial $11$ days time slots ($11*24*12 = 3168$ time slots) of traffic data are used during experiments. The initial $7$ days ($7*24*12 = 2016$ time slots) of traffic data in this time slot are used for training, and the subsequent $4$ days ($4*24*12 = 1152$ time slots) traffic data are used for testing. For G\'EANT, the three weeks of traffic data ($21*24*4 = 2016$ time slots) are used during experiments. The initial two weeks of traffic data ($14*24*4 = 1344$ time slots) are used for training of models, and the subsequent one week of traffic data ($7*24*4 = 672$ time slots) is used for testing. 
			
			\item The factorization rank $k$ is set to $20$ for all models.
			\item The models other than the proposed model are run with their default parameters.
			
			\item During training MCST-NMF, the penalty parameters $\lambda_{h}$ and $\lambda_{\mathcal{A}}$ for regularization terms are obtained using \eqref{eq:tune_lambdas} by setting $\beta_{h} = \beta_{\mathcal{A}}$. For Internet2, $\beta_{h} = \beta_{\mathcal{A}} = 0.2$. For G\'EANT, $\beta_{h} = 0.1$ and $\beta_{A} = 0.1$ respectively.
			
			\item During training MCST-NMF, the maximum number of iterations $q^{max}$ is set to $50$, while  $q_{W}^{max}$, $q_{H}^{max}$ and $q_{\Omega}^{max}$ are set to $10$. The parameter $\delta$ is set to $10^{-9}$,  $\delta_{W}$ and $\delta_{H}$ are set to $10^{-3}$, and   $\delta_{\Omega}$ is set to $10^{-5}$. 
	 		\item For testing MCST-NMF, Algorithm~\ref{algo:mcst_nmf_estimation} parameters are set as follows:
	 		\begin{inparaenum}[a)] 
	 			\item $\delta_{gd}$ and $\delta_{emi}$ are set to $10^{-3}$ and $10^{-9}$,  respectively,  and
	 			\item $q^{max}$ and $r^{max}$ are set to $200$. 
	 		\end{inparaenum}
			
		\end{enumerate}
		
	\paragraph{Choice of the lag set $\mathcal{L}$}\label{para:lag_set_choice} For training MCST-NMF using Algorithm~\ref{algo:mcst_nmf_training} requires a lag set. We strategically choose the lag set during experiments by keeping the general temporal behaviour of backbone computer network in our mind i.e. 
	\begin{inparaenum}[a)] 
 		\item network traffic behaviour persist for short period of time,
 		\item network traffic is likely to repeat its behaviour on a hourly basis,
 		\item network traffic may be similar after $8$ hours at the beginning and end of working hours and
 		\item network traffic exhibit diurnal pattern.
	\end{inparaenum}
	Based on the described strategy, the chosen lag sets are 
    \(
		\mathcal{L} = \{1, 2, 3, 12, 24, 96, 102, 108, 288\}
	\) 	
	and
	\(
    	\mathcal{L} = \{1, 4, 8, 32, 34, 36, 96\}
    \) 
    for Internet2 and G\'EANT respectively.
    
	All tests are preformed using MATLAB$^\circledR$ R2018b (Student License) under Windows 10$^\circledR$ environment on a laptop Intel$^\circledR$ CORE$^{\text{TM}}$ i5-3YY6U768Y0M CPU @2.60GHz 4GB RAM. The code of our proposed methods is available from  \url{https://github.com/5y3datif/MCST-NMF}. 

		\subsection{Experiments on Internet2}
		
		\begin{table}[htb]
			\caption{Statistical properties of the traffic estimation errors, SRE~\eqref{eq:SRE} and TRE~\eqref{eq:TRE}, on the test set for the Internet2 data set. 
			The lowest (best) values are highlighted in bold.} 
			\label{tb:abilene_sre_tre}
			\begin{subtable}[h]{\textwidth}
			\caption{Statistical properties of the SRE.}
			\label{subtb:abilene_sre}
			\centering
			\begin{tabular}{c c c c c c}
				\toprule[1.5pt]
				& {MCST-NMC} & {MCST-NMF} & {MNETME} & {CS-DME} & {PCA}\\
				\midrule[1.5pt]
				minimum & 0.07 & \textbf{0.05} & 0.07 & 0.81 & 0.07\\
				maximum & 1.36 & 1.34 & 11.86 & \textbf{1.03} & 2.37\\
				mean & 0.40 & \textbf{0.39} & 0.69 & 0.92 & 0.51\\
				median & \textbf{0.36} & \textbf{0.36}&	0.43 & 0.92 & 0.43\\			
				\makecell{standard \\ deviation} & 0.25 & {0.25} & 1.23 & \textbf{0.03} & 0.35\\
				\bottomrule[1.5pt]
			\end{tabular}
			\end{subtable}
			\newline
            \vspace*{0.5 cm}
            \newline
			\begin{subtable}[h]{\textwidth}
			\caption{Statistical properties of the TRE.}
			\label{subtb:abilene_tre}
			\centering
			\begin{tabular}{c c c c c c}
				\toprule[1.5pt]
				& {MCST-NMC} & {MCST-NMF} & {MNETME} & {CS-DME} & {PCA}\\
				\midrule[1.5pt]
				minimum & \textbf{0.05} & 0.06 & 0.13 & 0.80 & 0.08\\
				maximum & \textbf{0.30} & 0.31 & 0.37 & 0.97 & 0.54\\
				mean & \textbf{0.18} & \textbf{0.18} & 0.26 & 0.91 & 0.30\\
				median & \textbf{0.18} & 0.19 & 0.26 & 0.91 & 0.32\\			
				\makecell{standard \\ deviation} & 0.04 & 0.05 & 0.04 & \textbf{0.01} & 0.08\\
				\bottomrule[1.5pt]
			\end{tabular}
			\end{subtable}
		\end{table}
		
		Let us examine the results on the Internet2 dataset to assess the performance of our two proposed methods compared to the state of the art. 
		Table~\ref{tb:abilene_sre_tre} presents the statistical properties of SRE and TRE of the different methods.  
		
		SRE of our proposed methods, MCST-NMF and MCST-NMC, is the best in terms of the minimum, mean, median and standard deviation with values $0.05$, $0.39$, $0.36$ and $0.25$, respectively. The closest competitor in terms of the minimum, mean and median of SRE has values larger by $40\%$, $31\%$ and $19\%$, respectively. 
		In terms of the maximum value of SRE, our methods are the second best ($1.34$ and $1.36$). 
		Similarly, TRE of MCST-NMC and MCST-NMF have the lowest values w.r.t.\ all four criteria, namely the minimum ($0.05$ and$0.06$), maximum ($0.30$ and $0.31$), mean ($0.18$ and $0.18$) and median ($0.18$ and $0.19$). Its nearest competitor with respect to corresponding criteria has values larger by  $33\%$, $19\%$, $44\%$ and $44\%$, respectively. 
		
\begin{figure*}[ht!]
			\centering
			\begin{subfigure}[b]{0.48\textwidth}
			    \centering
				\includegraphics[width=\textwidth]{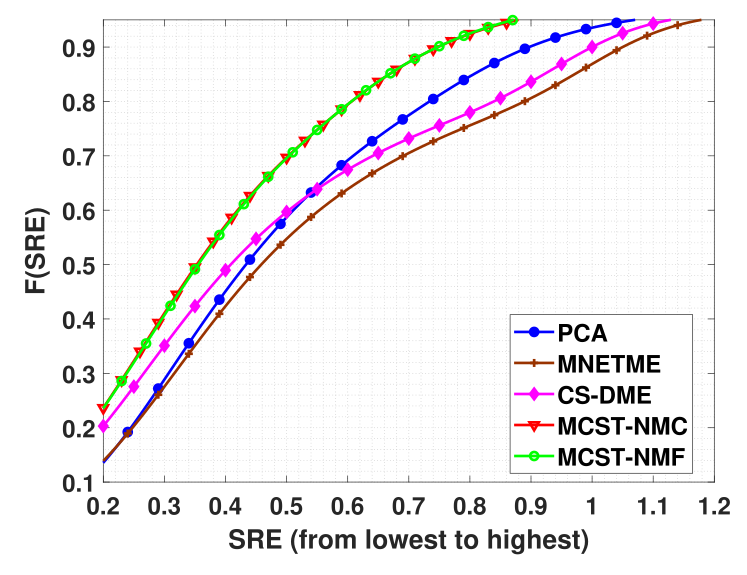}
			    \caption{CDFs of SRE}
			    \label{subfig:abilene_sre}
			\end{subfigure}
			\hfill
			\begin{subfigure}[b]{0.48\textwidth}
			    \centering
				\includegraphics[width=\textwidth]{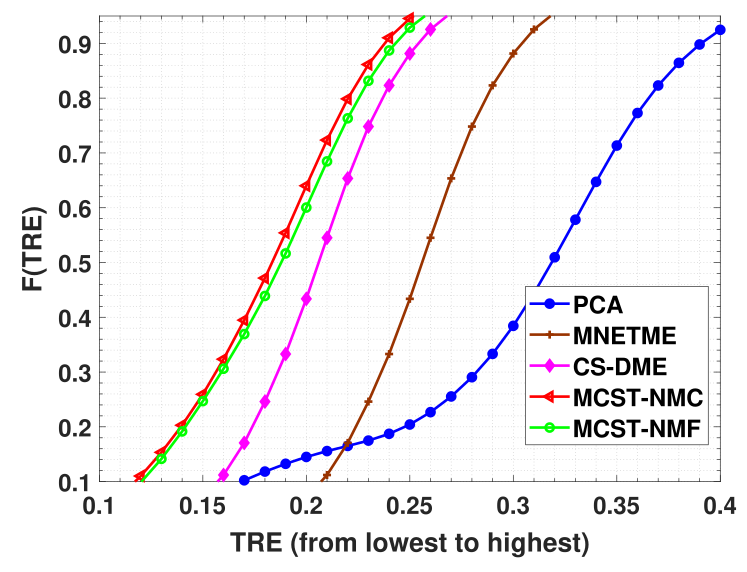}
			    \caption{CDFs of TRE}
			    \label{subfig:abilene_tre}
			\end{subfigure}
			\caption{Cumulative distributed function (CDF) of SRE (left) and TRE (right) resulting from comparing traffic estimation methods  over Internet2.}\label{fig:abilene_sre_tre}
		\end{figure*}

		To better visualize the differences, Figure~\ref{fig:abilene_sre_tre} displays the cumulative distribution functions (CDF) of the SRE and TRE of the different methods. 
		We observe that $90\%$ of OD flows estimated by our proposed methods (MCST-NMC and MCST-NMF) have SRE of at most $0.74$. In contrast, the nearest competitor, PCA, has SRE of at most $0.90$ for $90\%$ of OD flows that is $21.62\%$ higher comparing to that of MCST-NMF.  
		Results for the TRE for MCST-NMC and MCST-NMF are even more striking. All the models show low TRE but $90\%$ of OD flows estimated by our proposed methods (MCST-NMC and MCST-NMF) has TRE of at most $0.238$ and $0.242$, respectively, which is $6.7\%$ and $5\%$ smaller than the second best model, namely CS-DME (with value $0.254$) as shown on  Figure~\ref{subfig:abilene_tre}. 
		
		\subsubsection{Impact of the regularization parameters}

        This section analyzes the impact of the regularization parameters, $\beta_{h}$ and $\beta_{\mathcal{A}}$, on the performance of the proposed methods when tested over the Internet2 dataset. 
        To do so, we consider four cases by setting $\beta_{h}$ and $\beta_{\mathcal{A}}$ to their default value $0.5$ or to $0$, with a total of four cases. 
        All the remaining parameters are kept as described in the paragraph \nameref{para:parameter_setting_for_experiments} during all the conducted experiments. The obtained results are reported in Table~\ref{tb:abilene_sensitivity_analysis_regularization_parameters}.
                \begin{table*}[htb]
			\caption{Impact of the regularization parameters with respect to SRE and TRE for the Internet2 dataset. Lowest values are in bold.} \label{tb:abilene_sensitivity_analysis_regularization_parameters}
			\centering
			\resizebox{\textwidth}{!}{%
			\begin{tabular}{c c c c c}
				\toprule[1.5pt]
				& \multicolumn{2}{c}{SRE} & \multicolumn{2}{c}{TRE}\\
				\multirow{-2}{*}{ } & MCST-NMF & MCST-NMC & MCST-NMF & MCST-NMC\\
				\midrule[1.5pt]
				$\beta_{h}=0.5$, $\beta_{\mathcal{A}}=0.5$ & $\boldsymbol{0.73}$ & $\boldsymbol{0.73}$ & $0.187$ & $\boldsymbol{0.186}$\\
				$\beta_{h}=0.5$, $\beta_{\mathcal{A}}=0.0$ & $0.76$ & $0.76$ & $0.232$ & $0.230$\\
				$\beta_{h}=0.0$, $\beta_{\mathcal{A}}=0.5$ & $0.76$ & $0.77$ & $0.227$ & $0.216$\\
				$\beta_{h}=0.0$, $\beta_{\mathcal{A}}=0.0$ & $0.78$ & $0.78$ & $0.216$ & $0.216$\\
				\bottomrule[1.5pt]
			\end{tabular}
			}
		\end{table*}
		
        
        We observe that our two methods have the best performance when both the regularization terms are used. Using only one regularization term or none of the two regularization terms deteriorate SRE and TRE upto $24\%$ and $7\%$ respectively. This illustrates the effectiveness of the two regularization terms.
       Note that the case $\beta_{h} = 0$ represent the case where the lag set $\mathcal{L} = \emptyset$ which  has a significant impact on the TRE; see Table~\ref{tb:abilene_sensitivity_analysis_regularization_parameters}.

		\subsection{Experiment on G\'EANT}
		
		We now perform the same results as for Internet2 on the  G\'EANT data set.  
    	Table~\ref{tb:geant_sre_tre} presents the values of SRE and TRE for the different methods. 
    	\begin{table*}[htb]
			\caption{Statistical properties of SRE and TRE for PCA, MNETME, CS-DME and MCST-NMF over G\'EANT test data. Lowest values are in bold.} \label{tb:geant_sre_tre}
			\begin{subtable}[h]{\textwidth}
			\caption{Statistical properties of SRE}
			\label{subtb:geant_sre}
			\centering
			\begin{tabular}{c c c c c c}
				\toprule[1.5pt]
				& {MCST-NMC} & {MCST-NMF} & {MNETME} & {CS-DME} & {PCA}\\
				\midrule[1.5pt]
				minimum & \textbf{0.00} & \textbf{0.00} & \textbf{0.00} & 0.12 & 0.02 \\
				maximum & \textbf{23} & 46 & 59 & 30 & 37\\
				mean & 5.71 & 14.21 & 14.87 & \textbf{2.53} & 10.47 \\
				median & \textbf{0.82} & 0.86 & 0.87 & 1.27 & 0.91 \\			
				\makecell{standard\\ deviation} & 6.84 & 22.34 & 14.76 & \textbf{3.75} & 86.99\\
				\bottomrule[1.5pt]
			\end{tabular}
			\end{subtable}
			\newline
           \vspace*{0.5 cm}
            \newline
			\begin{subtable}[h]{\textwidth}
			\caption{Statistical properties of TRE}
			\label{subtb:geant_tre}
			\centering
			\begin{tabular}{c c c c c c}
				\toprule[1.5pt]
				& {MCST-NMC}  & {MCST-NMF} & {MNETME} & {CS-DME} & {PCA}\\
				\midrule[1.5pt]
				minimum & \textbf{0.05} & \textbf{0.05} & 0.06 & 0.41 & 0.06 \\
				maximum & \textbf{0.26} & \textbf{0.26} & 0.60 & 1.13 & 0.38\\
				mean & \textbf{0.09} & \textbf{0.09} & 0.13 & 0.64 & 0.12 \\
				median & \textbf{0.08} & \textbf{0.08} & 0.10 & 0.61 & 0.11 \\		
				\makecell{standard\\ deviation} & \textbf{0.03} & \textbf{0.03} & 0.10 & 0.12 & 0.05\\
				\bottomrule[1.5pt]
			\end{tabular}
			\end{subtable}
		\end{table*}

\begin{figure*}[ht!]
			\centering
			\begin{subfigure}[b]{0.48\textwidth}
			    \centering
			    \includegraphics[width=\textwidth]{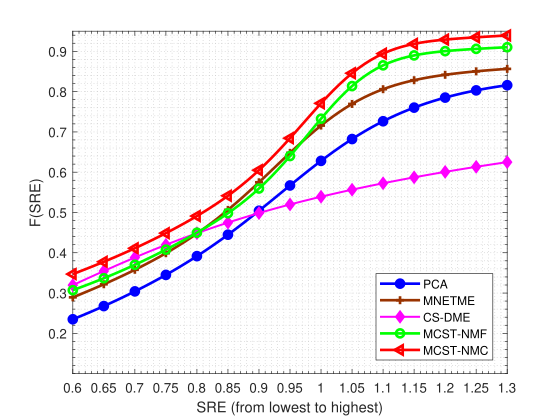}
			    \caption{CDFs of SRE}
			    \label{subfig:geant_sre}
			\end{subfigure}
			\hfill
			\begin{subfigure}[b]{0.48\textwidth}
			    \includegraphics[width=\textwidth]{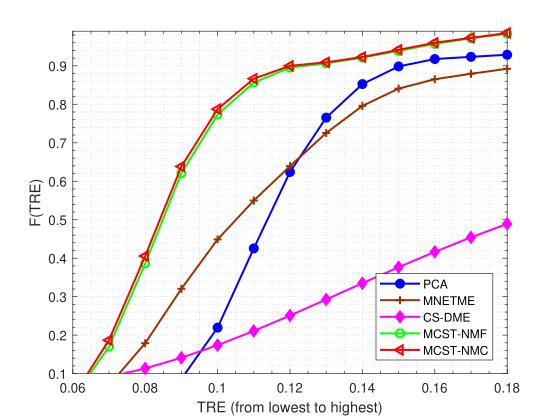}
			    \caption{CDFs of TRE}
			    \label{subfig:geant_tre}
			\end{subfigure}
			\caption{
			Cumulative distributed function (CDF) of SRE (left) and TRE (right) resulting from comparing traffic estimation models over G\'EANT test data. (On the right plot, the TRE of the two proposed variants, MCST-NMF and MCST-NMC,  overlap.)}
			\label{fig:geant_sre_tre}
		\end{figure*}

The SRE of our proposed methods is the best in terms of the median. 
    	Moreover, the matrix completion variant of our proposed model, MCST-NMC, is the best in terms of all criteria except except for the mean and standard deviation. The closest competitors of MCST-NMC in terms of the minimum, maximum and median of SRE has values larger by $0\%$, $30\%$ and $4.87\%$, respectively. The TREs of our proposed methods are the same. The closest competitor in terms of the minimum, maximum, mean, median and standard deviation has value larger by $20\%$, $46\%$, $33\%$, $25\%$ and $66\%$,  respectively. 
    	To further illustrate the differences, 
    Figure~\ref{fig:geant_sre_tre} shows the cumulative distribution functions (CDF) of the different methods. 
    We observe that $80\%$ of OD flows estimated by our proposed methods, MCST-NMC and MCST-NMF, have SRE of value at most $1.02$ and $1.04$, respectively. 
    The second best performing model (MNETME) has SRE of at most $1.1$ for $80\%$ of OD flows. This value is larger by $8\%$ compared to MCST-NMC;  see Figure~\ref{subfig:geant_sre}. 
    Results of TRE for MCST-NMC and MCST-NMF are very similar. 
    All the models show low TRE but $90\%$ of OD flows estimated by MCST-NMC and MCST-NMF have TRE of at most $0.12$ whereas the second best model, PCA,  has TRE of at most $0.15$ for $90\%$ of OD flows. This value is larger by $25\%$; see Figure~\ref{subfig:geant_tre}. 
    	 
        \subsubsection{Impact of the regularization parameters}

        This section analyzes the impact of the regularization parameters, $\beta_{h}$ and $\beta_{\mathcal{A}}$ on the performance of proposed solutions when tested over G\'EANT dataset, exactly as for the Internet2 dataset.       
    Table~\ref{tb:geant_sensitivity_analysis_regularization_parameters} reports the values of the SRE and TRE for the different values of the regularization parameters.         
          \begin{table*}[htb]
			\caption{Impact of the regularization parameters with respect to SRE and TRE over G\'EANT. Lowest values are in bold.} \label{tb:geant_sensitivity_analysis_regularization_parameters}
			\centering
			\resizebox{\textwidth}{!}{%
			\begin{tabular}{c c c c c}
				\toprule[1.5pt]
				& \multicolumn{2}{c}{SRE} & \multicolumn{2}{c}{TRE}\\
				\multirow{-2}{*}{cases} & MCST-NMF & MCST-NMC & MCST-NMF & MCST-NMC\\
				\midrule[1.5pt]
				$\beta_{h}=0.5$, $\beta_{\mathcal{A}}=0.5$ & $1.20$ & $\boldsymbol{1.12}$ & $0.124$ & $\boldsymbol{0.121}$\\
				$\beta_{h}=0.5$, $\beta_{\mathcal{A}}=0.0$ & $1.27$ & $1.21$ & $0.125$ & $0.123$\\
				$\beta_{h}=0.0$, $\beta_{\mathcal{A}}=0.5$ & $4.68$ & $1.70$ & $0.124$ & $\boldsymbol{0.121}$\\
				$\beta_{h}=0.0$, $\beta_{\mathcal{A}}=0.0$ & $4.25$ & $1.80$ & $0.124$ & $\boldsymbol{0.121}$\\
				\bottomrule[1.5pt]
			\end{tabular}
			}
		\end{table*}
        
        
        We observe that both proposed methods exhibit the best performance when  the regularization terms are used. This is particularly true for the SRE, with a significant deterioration for both methods; namely from 1.20 to 4.25 for MCST-NMF, and from 1.12 to 1.80 for MCST-NMC. 
        Moreover, we observe that the use of the lag set is particularly important to have low SRE values, as using $\beta_{h} = 0$ significantly impacts the SRE. 

	\section{Conclusion}\label{sec:conclusion}

	In this paper, we have proposed an NMF-based approach to tackle the network traffic flow estimation problem. 
	To the best of our knowledge, it is the first time NMF is used for this specific task.   
	A notable shortcoming of previously explored dimensionality-reduction approaches (e.g., based on PCA) was to solve these problems by ignoring the nonnegativity constraints.  	
	Moreover, our approach uses two regularizers, namely orthogonality to better cluster the data, and autoregression to take the temporal correlations into account, which further improves its performance. 
	
	We proposed their two different variants of our model: 
	(1)~MCST-NMF that is most suitable when the training dataset does not have missing entries, and (2) MCST-NMC that is designed specifically to handle missing entries.   
	We have shown on two  real-world data sets, namely G\'EANT and Internet2, that	our methods outperform existing techniques based on linear dimensionality reduction.

	\paragraph{Further Works}	
	Recently, deep matrix factorization models have emerged; in particular {deep NMF} has been shown to be able to  capture several layers of meaningful features; see,  e.g.,~\cite{huang2019self,huang2020regularized,dedeep}. 
	It would be interesting to adapt our NMF based traffic matrix estimation model into a deep NMF model. 
	Other directions of research include the use of other NMF models to perform the traffic flow estimation, such as~\cite{xu2021learning,nie2018spatio}, or to improve prediction using data coming from other sources than the traffic flows and then use multi-view techniques such as~\cite{kumar2020multi,huang2020auto}.

\small
	\section*{Acknowledgement} 
	The authors would like to thank the reviewers for the insightful feedback that helped improve the paper significantly. 
	
		The financial support of HEC Pakistan is highly acknowledged for granting PhD scholarship to the first author. 
		 NG acknowledges the support by the ERC starting grant No 679515, and by the Fonds de la Recherche
Scientifique - FNRS and the Fonds Wetenschappelijk Onderzoek - Vlaanderen (FWO) under EOS project O005318FRG47.



\bibliographystyle{elsarticle-num} 
\bibliography{references}






\section*{CRediT author statement}
\textbf{Syed Muhammad Atif:} Conceptualization, Methodology, Software, Validation, Formal analysis, Investigation, Visualization, Investigation, Writing-Original draft preparation. \textbf{Nicolas Gillis:} Supervision, Visualization, Writing-Original draft preparation, Writing- Reviewing and Editing. \textbf{Sameer Qazi:} Supervision, Visualization, Writing- Reviewing and Editing. \textbf{Imran Naseem:} Supervision, Visualization, Writing- Reviewing and Editing.
\section*{}
\begin{wrapfigure}{l}{1.1in}
    \centering
    \includegraphics[width=0.75in,height=1.25in,clip,keepaspectratio]{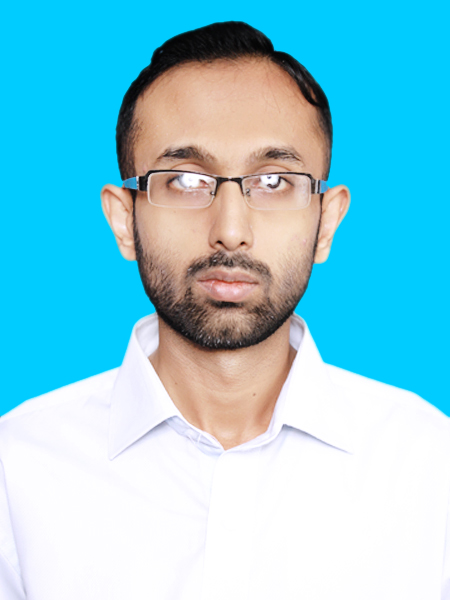}
\end{wrapfigure}

\textbf{Syed Muhammad Atif} completed his Bachelors in Computer Engineering from Sir Syed University of Engineering and Technology. He completed his MS in Computer Networks from Usman Institute of Technology. He is currently pursuing a PhD in Computer Science from PAF-KIET under HEC Indigenous Fellowship Program. He is the author or co-author of 12 publications those were published in reputable journals. His current areas of research are Non Negative Matrix Factorization, Deep Learning, Sentiment Analysis, Network Tomography and Green Routing.

\section*{}
\begin{wrapfigure}{l}{1.1in}
    \centering
    \includegraphics[width=0.9in,height=1.25in,clip,keepaspectratio]{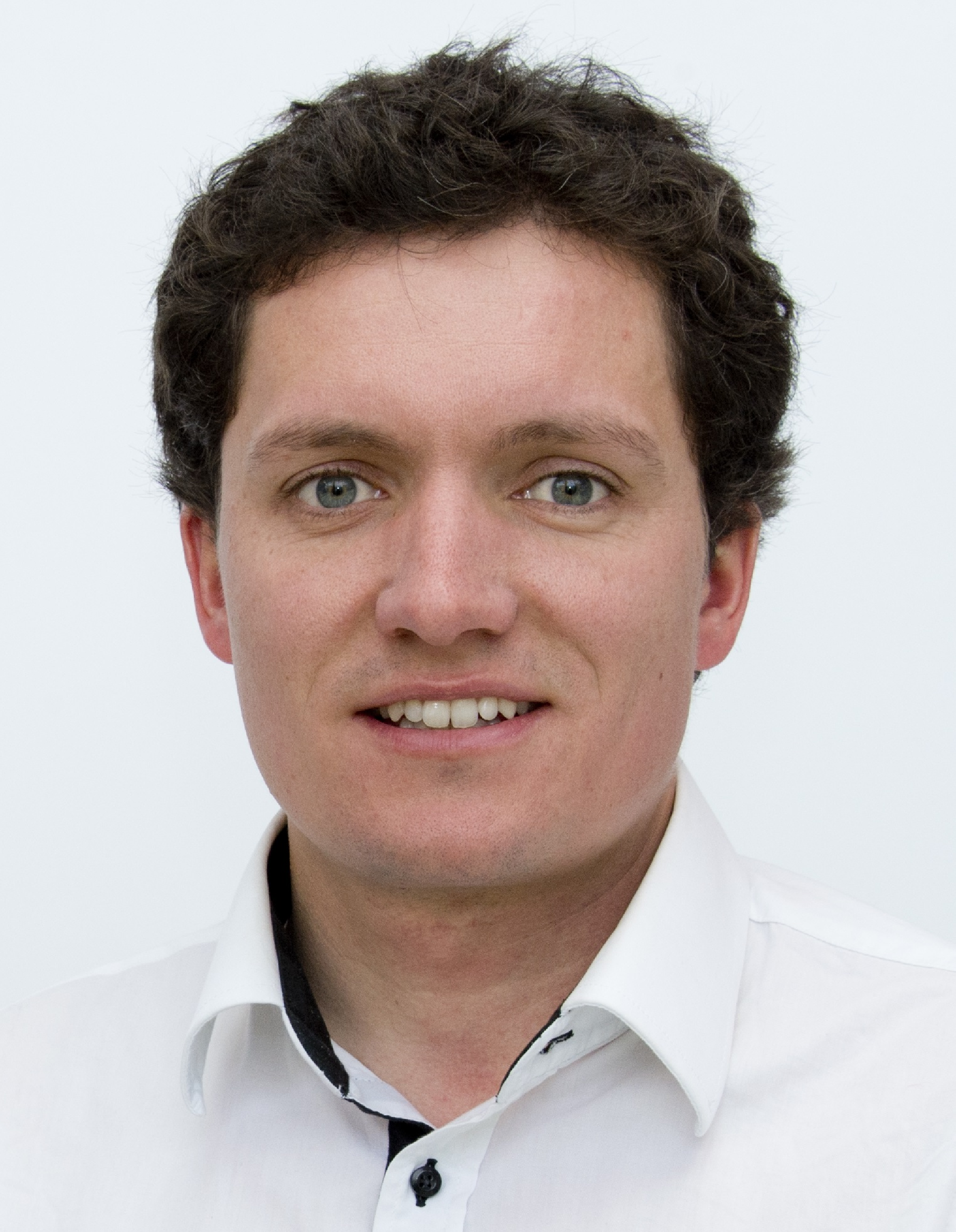}
\end{wrapfigure}

\textbf{Nicolas Gillis} received his M.Sc. and Ph.D. degrees in applied mathematics from UCLouvain, Belgium, in 2007 and 2011, respectively. He is currently an associate professor in the Department of Mathematics and Operational Research, University of Mons, Belgium. His research interests include optimization, numerical linear algebra, signal processing, machine learning, and data mining. He received the Householder Award in 2014 and a European Research Council starting grant in 2015. 
He has published a book on `Nonnegative Matrix Factorization' in 2020 with SIAM. 
He currently serves as an associate editor of IEEE Transactions on Signal Processing and SIAM Journal on Matrix Analysis and Applications.

\section*{}
\begin{wrapfigure}{l}{1.1in}
    \centering
    \includegraphics[width=1in,height=1.25in,clip,keepaspectratio]{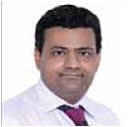}
\end{wrapfigure}

\textbf{Sameer Qazi} received his B.E. degree from National University of Sciences and Technology, Pakistan, in 2001 and the MS and PhD degrees from the University of New South Wales, Australia, in 2004 and 2009. He is currently working as Associate Professor in the College of Engineering at Karachi Institute of Economics and Technology, Pakistan. His research interests are Computer Network Optimization Problems, Network Tomography, UAV based video surveillance applications and Cloud Computing.

\section*{}
\begin{wrapfigure}{l}{1.1in}
    \centering
    \includegraphics[width=1in,height=1.25in,clip,keepaspectratio]{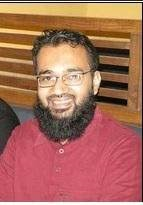}
\end{wrapfigure}

\textbf{Imran Naseem} received the BE degree in electrical engineering from the NED University of Engineering and Technology, Pakistan, in 2002, the MS degree in electrical engineering from the King Fahd University of Petroleum and Minerals (KFUPM), KSA, in 2005, and PhD degree from the University of Western Australia, in 2010. He did his post doctorate in the Institute for Multi-sensor Processing and Content Analysis, Curtin University of Technology, Australia. He joined the College of Engineering, KIET, Pakistan, in 2011 where he is currently an associate professor. He is also an adjunct research fellow in the School of Electrical, Electronic, and Computer Engineering, University of Western Australia. His research interests include pattern classification and machine learning with a special emphasis on biometrics and bioinformatics applications. He has authored several publications in top journals and conferences including the IEEE Transactions on Pattern Analysis and Machine Intelligence, IEEE International Conference on Image Processing etc. His benchmark work on face recognition has received more than 180 citations in less than four years. He is also a reviewer of the IEEE Transactions on Pattern Analysis and Machine Intelligence, the IEEE Transactions on Image Processing, and the IEEE Signal Processing Letters.

\end{document}